\documentstyle[11pt,rtrpp4]{article}

\tighten

\received{}
\accepted{}
\journalid{}{}
\articleid{}{}
\righthead{FERRARO, ET AL.}
\lefthead{IR SURVEY  OF GLOBULAR CLUSTERS}
 
\slugcomment{In press in The Astronomical Journal}
 
\begin{document}
 
\title{
A new IR--Array photometric survey of Galactic Globular Clusters:
A detailed study of the RGB sequence as a step towards
the global testing of stellar models.
\footnote{
Based on data taken at the ESO--MPI 2.2m Telescope equipped
with the near IR camera IRAC2 - ESO, La Silla (Chile).}
 }
\author{Francesco R. Ferraro}
\vskip2truemm
\affil{European Southern Observatory, Karl Schwarzschild Str. 2,
        D--85748 Garching bei M\"unchen, Germany\\
    On leave from Osservatorio Astronomico di Bologna\\
   e--mail: fferraro@eso.org, ferraro@bo.astro.it}
 
\and 

\author{Paolo Montegriffo, Livia Origlia}
\vskip2truemm
\affil{Osservatorio Astronomico di Bologna, 
        Via Ranzani 1, I--40127 Bologna, Italy\\
   e--mail: pmontegr,origlia@bo.astro.it}
 
\and 

\author{Flavio Fusi Pecci}   
\vskip2truemm
\affil{ Stazione Astronomica, 09012 Capoterra, Cagliari, Italy\\
On leave from Osservatorio Astronomico
       di Bologna.\\
   e--mail: flavio@bo.astro.it}
 
\begin{abstract}
      
We present high quality near infrared Color Magnitude Diagrams of 10
Galactic Globular Clusters (GCs) spanning a wide metallicity range
($-2.15<[Fe/H]<-0.2$).
This homogeneous data--base has been used to perform a detailed
analysis of the Red Giant Branch (RGB), adopting a variety of observables
to describe its physical and chemical properties.

First, a set of  metallicity indicators have been measured, namely:
{\it (i)} the RGB (J--K) and (V--K) colors at different magnitude levels;
{\it (ii)} the RGB K magnitude at different colors;
{\it (iii)} the RGB slope.
For these parameters we present new calibrations in terms of both
spectroscopic iron abundance and {\it global} metallicity, including
the $\alpha$--element enhancement. These relations can be used to derive
a {\it photometric} estimate of the GC metal content from the RGB
morphology and location.

Second, the location in luminosity of the main RGB features (namely,
the Bump and the Tip) and their dependence on metallicity have been
studied, yielding quantitative observational relationships.

Finally, adopting new transformations between the observational and
theoretical quantities, the mean ridge lines for the clusters of our sample
have been reported in the $(M_{Bol}, Log(T_e))$ plane.
This allows to study the RGB location in terms of effective temperature, 
bolometric luminosity of the main RGB features, and their calibrations
with varying metallicity.
Direct comparisons between up--dated theoretical models and observations 
show an excellent overall agreement.

\end{abstract}
 
\keywords{techniques: photometric --- surveys     
          --- stars: fundamental parameters -- stars: late--type -- 
          stars: Population II}
 
\section{Introduction}
 
Stellar evolution theory is crucial to yield a reliable clock
for dating astrophysical objects. Suitable  Color Magnitude Diagrams
(CMDs) and Luminosity Functions (LFs) are the most powerful tools to test
theoretical models and, in turn, the {\it running} of the stellar clock.
Within this framework, our group started a long--term project devoted to the
quantitative analysis and testing of each individual evolutionary sequence
in the CMDs of Galactic Globular Clusters (GGCs).

In this paper we present the results of a detailed study of the
Red Giant Branch (RGB), using CMDs and LFs in the near--IR bands.
Since the contrast between the red giants and the unresolved background
population in the IR bands is greater than in any optical region,
they can be observed with the highest S/N ratio also in the innermost
region of the cluster. Moreover, when combined with optical observations,
IR magnitudes provide useful observables such as for example the
V--K color, which is an excellent indicator of the stellar effective
temperature (T$_{e}$).

The advantage of observing GGCs in the near IR is well known since many
years. The first systematic IR survey of GGCs was carried out in their
pioneering work by Frogel, Cohen \& Persson (1983, hereafter FCP83), which
provided in particular the first quantitative description of the RGB
location with varying the cluster metal abundance.
However, due to intrinsic technical limits of the old single--channel detectors
and aperture photometry, the FCP83  data--base includes only a few bright
stars in the external regions of the clusters (a total of $\sim $350 stars
in 30 GGCs).
A detailed comparison with the theoretical models based on suitable LFs was
thus actually impossible because of the small number of observed stars.

The advent of new IR arrays with pixel sizes and overall performances close
to those of optical CCDs has then opened new perspectives to
the construction of large and complete samples of RGB stars with high
photometric accuracy (Davidge \& Simons 1991, 1994a,b; Minniti 1995,
Minniti et al. 1995, Ferraro et al. 1994a,b, Ferraro et al. 1995, Montegriffo
et al. 1995, Kuchinski et al. 1995--hereafter K95, Kuchinski \& Frogel
1995--hereafter KF95, Guarnieri et al. 1998).

The main aim of our project is to obtain a complete quantitative description
of the RGB as a function of the intrinsic cluster parameters, and to
yield a few observational relationships for general use and suitable
to carefully test the theoretical models.
Schematically, we study:

{\it i)~}the location in color of the RGB, its morphology and slope,
and their dependence on metal abundance.

{\it ii)~} the extent in luminosity of the RGB; this provides a quantitative
determination of the stellar luminosity at the helium-flash (the RGB--tip)
with varying metallicity, setting also strong constraints on mass loss
and the calibration of a useful method to determine the cluster distance
scale.

{\it iii)~}the precise location in luminosity and temperature of the
so--called RGB bump and of any other peculiarity (gaps, clumps, etc.)
one might detect in the available data.

The calibrations eventually obtained depend on the adopted reddening,
metallicity and distance scale, and on the assumed transformations
between the observed and theoretical planes (especially the bolometric
corrections and the temperature scale).

In Sect.2 we present the IR data of the selected sample of GGCs.
A wider presentation of some individual clusters can be found in
Ferraro et al. (1994a), Montegriffo et al. (1995) and Guarnieri et al.
(1998).
In Sect.3 we describe the procedure used to compute the fiducial RGB
ridge lines.
In Sect.4 we present the basic assumptions on metallicity scales, 
distance modulus 
and reddening adopted in this paper.
In Sect.5 we analyze the general RGB location and morphology in the 
observational
planes.
In Sect.6 we present our results on the observed RGB features 
and the comparison with previous work and up--dated theoretical models.
In Sect.7 we discuss the transformations to the theoretical plane 
and in Sect.8 we summarize our conclusions.

\begin{figure*}[htb]
\vskip6.9truein
\includegraphics{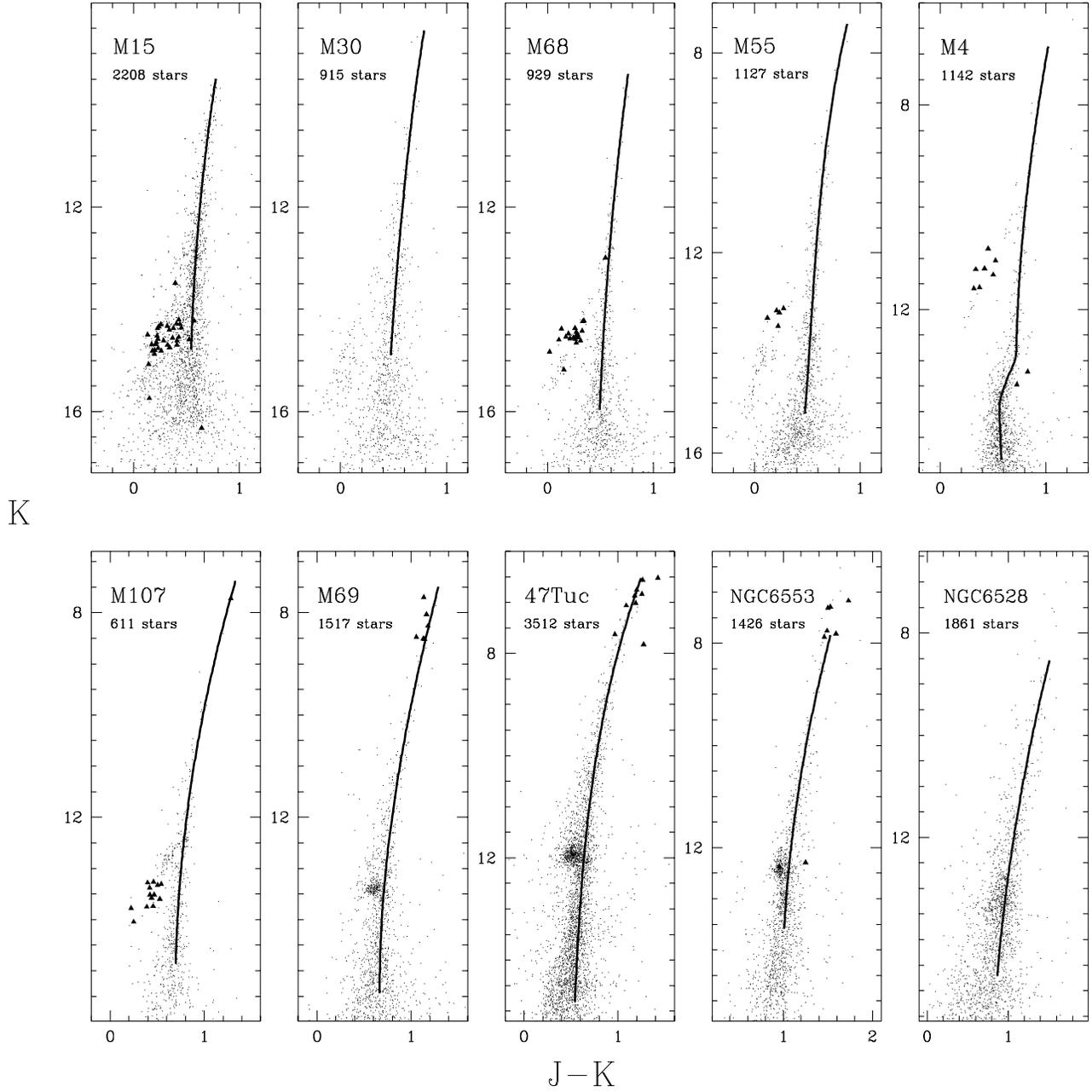}
\figcaption[]{
(K,J--K) Color--magnitude diagrams for the 10
GGCs in our data--base.
The thick line in each panel indicates the fiducial ridge line
of the RGB.
Variable stars have been plotted as filled triangles. 
 }
\end{figure*}

\section{The Sample}

The IR data set of globular clusters presented here was obtained
at ESO, La Silla (Chile), during two different runs (on June 1992 and
June 1993), using the ESO--MPI 2.2m telescope and the near IR camera
IRAC--2 (Moorwood et al. 1992) equipped with a NICMOS--3 256x256 array
detector.
The frames were taken through standard J and K broad-band filters.
For each cluster we mapped the central $4\times 4$ square arcminutes
using two different magnifications:
$0.27"$/pixel in the central field, and $0.5"$/pixel for the four
partially overlapping fields centered at $\sim $100 arcsec NE,
NW, SW and SE, respectively, from the cluster center.
In the case of M4 only the central high spatial resolution frames were used.

The resulting images were averages of typically 60 exposures of 1 sec
integration time each, and were sky subtracted and flat-field corrected.
The sky was monitored every 2 minutes at a distance of about 300 arcsec from
the cluster center.

More details on the pre--reduction procedure and the photometric calibration
can be found in Ferraro et al. (1994a) and Montegriffo et al. (1995).
Here we just remark that most of the photometric reductions were carried
out using the ROMAFOT package (Buonanno et al. 1983), specifically adapted
to work with undersampled stellar images (Buonanno \& Iannicola, 1988).

The calibration curves linking the aperture photometry to the standard
system are reported in Ferraro et al. (1994a) and Montegriffo et al. (1995)
for the June 1992 and June 1993 runs, respectively.
During these two runs we secured images of a sample of 15 GGCs but here we
present the results for the 10 GGCs (namely M15, M30, M68, M55, M4, M107,
M69, 47 Tuc, NGC6553, NGC6528) with the best quality data: the total
sample contains J and K photometry of about 17000 stars.
The V magnitude used to derive the V--K color is taken from 
recent published photometry, according to Table 1 by Montegriffo et al. (1998).

\begin{figure*}[htb]
\vskip6.9truein
\includegraphics{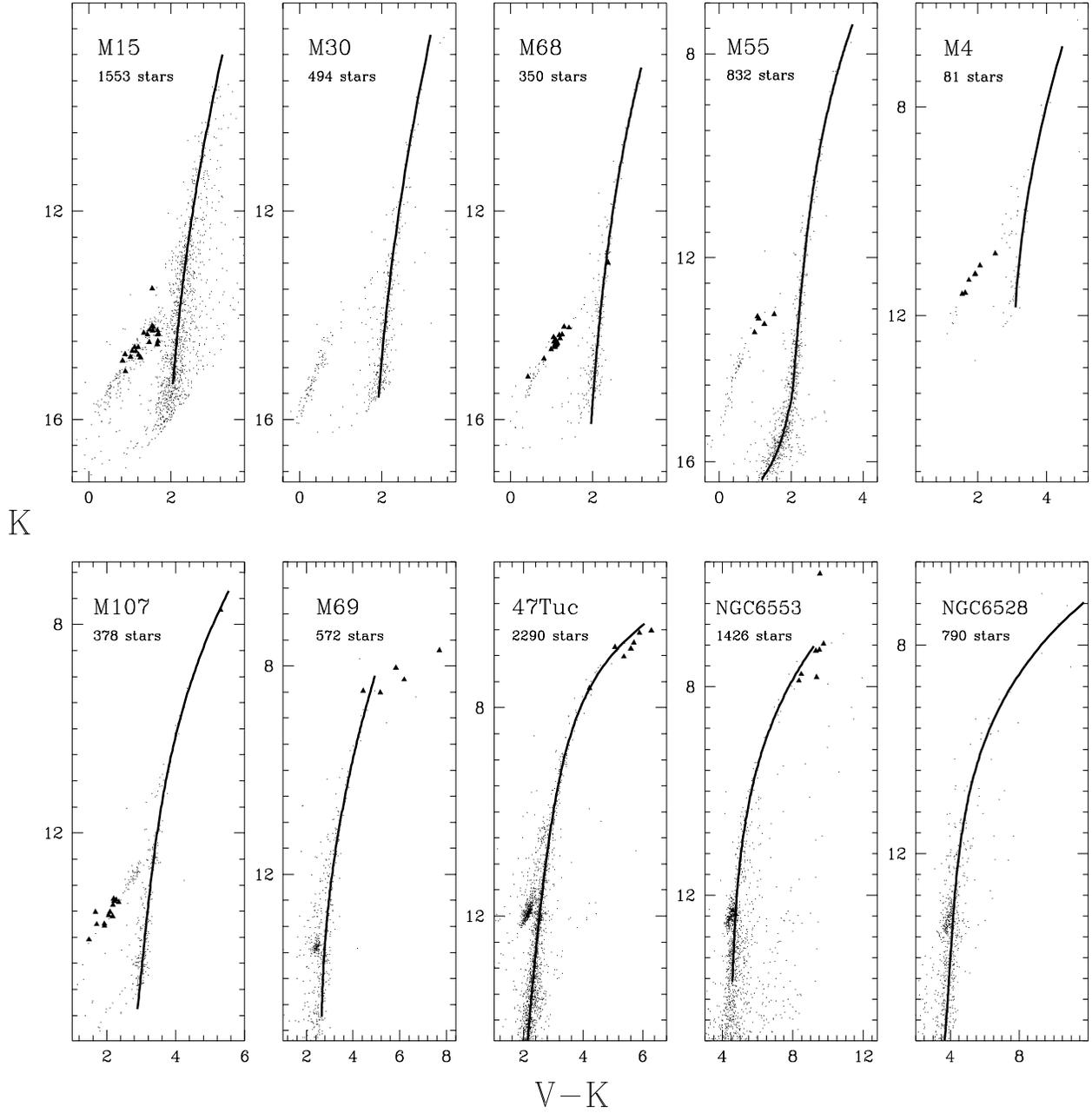}
\figcaption[]{
(K,V--K) Color--magnitude diagrams for the 10
GGCs in our data--base.
The thick line in each panel indicates the fiducial ridge line of the
RGB.
Variable stars have been plotted as filled triangles.
}
\end{figure*}

\section{The RGB fiducial ridge lines}

In order to compute the fiducial RGB ridge lines for the clusters in
our sample  we followed a standard procedure already used by other
authors (Da Costa \& Armandroff 1990; Ferraro et al. 1999, hereafter F99).

A first rough selection of the candidate RGB stars was performed on the
CMDs by eye, removing the HB and part of the AGB stars.
This operation is relatively easy for all the clusters of our sample
(cf. Fig.1,2), except the most metal rich ones, like NGC6528, where the
HB merges the RGB without displaying any discontinuity.

As a second step, we used a low order polynomial to fit the selected stars 
and we rejected those at $\ge \pm 2\sigma $ from the best fit line.
The procedure was iterated until the best stable fit to the overall shape
of the RGB was obtained.

The final ridge lines overplotted to the (K,J--K) and (K,V--K) CMDs
are shown in Fig.1 and Fig.2, respectively.

\begin{deluxetable}{llcccccc}
\footnotesize
\tablewidth{16truecm}
\tablecaption{Adopted parameters for the observed GGCs.}
\tablehead{
\colhead{Cluster} &
\colhead{} &
\colhead{$[Fe/H]_{Z85}$} &
\colhead{$[Fe/H]_{CG97}$} &
\colhead{$[M/H]$} &
\colhead{$E(B-V)$} &
\colhead{$V(ZAHB)$} &
\colhead{$(m-M)_0$}
}
\startdata
 NGC104  & 47Tuc & --0.71 & --0.70 & --0.59 & $0.04\pm0.02$ & $14.22\pm0.07$ & 13.32 \nl
 NGC4590 & M68   & --2.09 & --1.99 & --1.81 & $0.04\pm0.02$ & $1
5.75\pm0.05$ & 15.14 \nl
 NGC6121 & M4    & --1.33 & --1.19 & --0.94 & $0.36\pm0.05$ & $1
3.45\pm0.10$ & 11.68 \nl
 NGC6171 & M107  & --0.99 & --0.87 & --0.70 & $0.33\pm0.05$ & $1
5.72\pm0.10$ & 13.95 \nl
 NGC6528 &       & --0.23 & --0.38 & --0.31 & $0.62\pm0.10$ & $1
7.17\pm0.20$ & 14.37 \nl
 NGC6553 &       & --0.29 & --0.44 & --0.36 & $0.84\pm0.10$ & $1
6.92\pm0.20$ & 13.46 \nl
 NGC6637 & M69   & --0.59 & --0.68 & --0.55 & $0.17\pm0.04$ & $1
5.95\pm0.10$ & 14.64 \nl
 NGC6809 & M55   & --1.81 & --1.61 & --1.41 & $0.07\pm0.04$ & $1
4.60\pm0.10$ & 13.82 \nl
 NGC7078 & M15   & --2.17 & --2.12 & --1.91 & $0.09\pm0.04$ & $1
5.90\pm0.07$ & 15.15 \nl
 NGC7099 & M30   & --2.13 & --1.91 & --1.71 & $0.03\pm0.02$ & $1
5.30\pm0.10$ & 14.71 \nl
\enddata
\tablecomments{Global metallicity ($[M/H]$), $V(ZAHB)$ and $(m-M
)_0$
are from Ferraro et al. (1999).}
\end{deluxetable}

\section{Basic assumptions}

In this Section 
we discuss the basic assumptions and the corresponding uncertainties 
on the adopted metallicity scales, distance moduli and reddening
for the clusters in our sample 
and their impact on the determination of the 
RGB parameters.

\subsection{The cluster metallicity}

The metallicity measurements most frequently used for GGCs refer to
the classical scale proposed by Zinn and collaborators in the 80's
(Zinn 1980, Zinn \& West 1984, Zinn 1985, hereafter Z85). Even though
new homogeneous observations of metallicity  indicators have been obtained
(cf. e.g. Carretta \& Gratton, 1997 -- hereafter CG97 --  Rutledge,
Hesser \& Stetson 1997), the Zinn's scale still represents the
most complete data--set available in the literature. There are however
reasons, discussed in the recent quoted papers, which suggest to
slightly revise the Zinn's scale according to the latest high--resolution
spectral data.

In particular, CG97 have recently presented high quality measurements of
iron abundances using high dispersion spectra of $FeI$ and $FeII$ lines.
They observed stars in a sample of 24 GGCs spanning the metallicity range
$-2.24<[Fe/H]<-0.5$. By comparing their measurements with those published
by Z85, they noted that there are systematic differences (up to $\delta
[Fe/H]\sim 0.2$) with respect to the Z85 scale, especially for intermediate
metallicities ([Fe/H]$\sim-1.5$).
CG97 derived a quadratic relation  to transform the Z85 data in their
own scale (cf. their eq. 7), which we will adopt in the following
as reference. Consequently, since only 5 out of 10 GGCs considered in this
paper have direct measurements in the GC97 list, for the other 5
the metallicity has been obtained by transforming the Z85 estimate into the
GC97 scale (cf. F99).

A further consideration is important: in order to perform a correct
parametrization of the RGB behavior as a function of metallicity,
the simple knowledge of the quantity usually called {\it iron abundance}
is not sufficient. In fact, it is known since many years (Renzini 1977) that
the scaling of the RGB location in the CMD with metallicity is essentially
due to the changing of the stellar opacity.

The main source of the continuum opacity in the temperature range
3000--6000K, typical of the old RGB population of GGCs, are the $H^-$ ions.
They form by capturing the electrons provided by metals with a
low--ionization potential (mainly Mg, Si, Fe). For this reason, as noted
among others by  Straniero \& Chieffi  (1991), and  Salaris \&
Cassisi (1996), the RGB location mainly depends on the [Mg+Si+Fe]
mixture abundance rather than on the [Fe] abundance alone. Therefore, a
more reliable parameter to describe the actual metal content of the RGB
stars is the so--called {\it global} metallicity, which takes into account
not only the iron but also the the
$\alpha$--element (like Mg and Si) abundance.

There are many observational evidences (cf. e.g. Carney 1996 and references
therein) that $\alpha$--elements are enhanced with respect to iron
in Population II stars. While there is a quite general consensus on the
size of the mean overabundance ([$\alpha$/Fe]$\sim$0.3 dex), the actual
trend with metallicity is not firmly established (cf. Carney 1996 and
Origlia et al. 1997 for references).

\begin{figure*}[htb]
\vskip6.9truein
\includegraphics{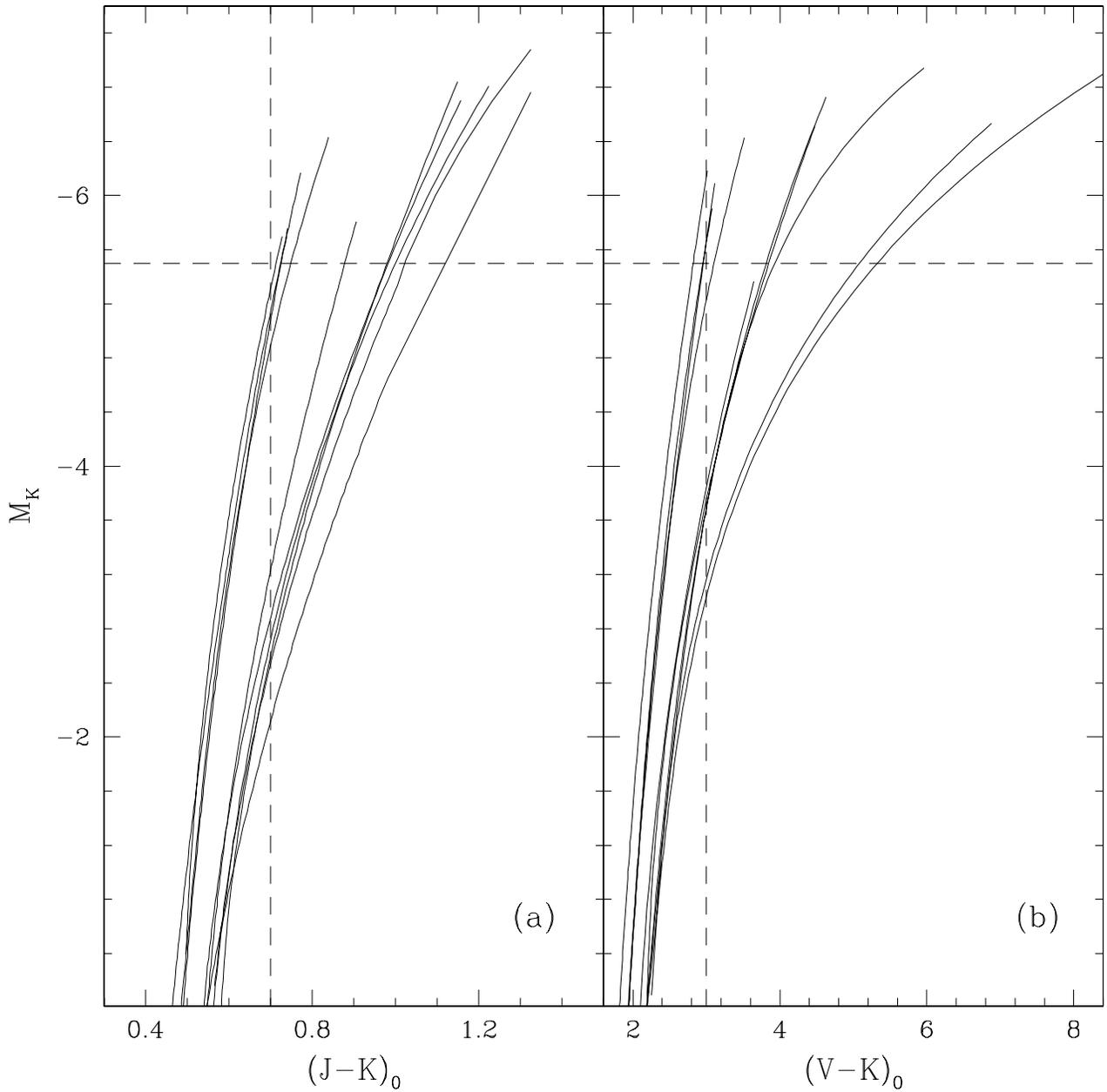}
\caption{
RGB fiducial ridge lines for the 10 GGCs in our sample in the
M$_K$,(J--K)$_0$  and
M$_K$,(V--K)$_0$ planes, ({\it panel (a)} and {\it  (b)}), respectively.
The dashed lines indicate the magnitude levels at which some of the
parameters defined in the text are read.
}
\end{figure*}

Straniero \& Chieffi  (1991) and Salaris, Chieffi \& Straniero (1993)
demonstrated that, when computing the isochrones of Population II stars,
the contribution of the $\alpha$--element enhancement can be taken
into account by simply re--scaling standard models to the {\it global}
metallicity [M/H] according to the following relation:
$$ [M/H] = [Fe/H] + log_{10} (0.638 f_{\alpha}+0.362)$$
where $f_{\alpha}$ is the enhanced factor of the $\alpha$--elements.

Following these prescriptions, in F99 we computed the {\it global}
metallicity for a sample of 61 GGCs within the framework of a
systematic study of the evolved sequences in the (V,B--V) CMD.
The values assumed here and listed in Table 1 have been thus taken from
Table 2 in F99. Schematically, in F99 we assumed a constant
$[\alpha/Fe]=0.28$ for $[Fe/H]<-1$ and linearly decreasing to zero
for $-1<[Fe/H]<0$. We also carefully discussed there the possible effect
of different assumptions in the $\alpha-$enhancing relation. In particular,
we showed that, adopting a constant enhancement over the
entire range of metallicity ($-2<[Fe/H]<0$) (Carney 1996) rather than
a linear decreasing trend for $[Fe/H]>-1$, 
does not produce any significative difference
in the description of the main RGB features in the (V,B--V) plane.

\subsection{The distance scale and reddening}

To convert the RGB fiducial lines into the absolute planes it is necessary
to adopt a distance scale and to correct for reddening.

Despite the huge observational efforts, the definition of the most
suitable distance scale for GGCs is still very controversial and
beyond the purposes of the present study (see for discussion and
references VandenBerg et al. 1996, Chaboyer et al.
1996, Gratton et al. 1997, Caloi et al. 1997, Reid 1998,
Pont et al. 1998, Carretta et al. 1999).

In the following we adopt the distance scale discussed
and eventually established in F99. In that paper  a new methodology to derive 
the actual level of the ZAHB and the distance moduli from the matching
of V(ZAHB) and the theoretical models computed by Straniero, Chieffi 
\& Limongi (1997, hereafter SCL97) has been presented.

The theoretical
relation adopted by F99 for the ZAHB is the following:
$$M_V^{ZAHB}= 0.0458 [Fe/H]^2 +0.3485 [Fe/H]+1.0005$$
\noindent 
which, in the range $-2.2<[M/H]<-0.4$, can also be described
by the linear best--fit regression:
$$M_V^{ZAHB}=0.23([Fe/H]+1.5)+0.595 $$

\noindent 
Among the many possible choices within the range of the so--called
{\it long} and {\it short} distances scales (see references above),
this relation yields intermediate values for the GGC distance moduli.

The reddening corrections have been computed using
the latest compilation by Harris (1996).
In Table 1, 
the final metallicity (cf. Sect.4), reddening (Harris 1996) 
and distance modulus (from F99, Table 2, column 7)
adopted in this paper for the 10 GGCs in our sample are reported.

The impact of different assumptions for the cluster distance scale
and reddening will be discussed in Sect. 6.1.1 and 6.1.2.

\subsection {Errors}

Though the uncertainty in the determination of the observable $V(ZAHB)$ is 
$\sim 0.1$ mag (see column 6 in Table 1) for most of the clusters in our 
sample, 
according to F99 we conservatively
estimate that the {\it global} uncertainty on the
derived distance moduli listed in Table 1 is of the order of $\sim 0.2$ mag.
This estimate takes also into account 
the effect of the errors  in the adopted 
metal abundances, reddening etc..
The distance modulus uncertainty 
turns to be the driving factor in the error
budget of most of the RGB parameters derived in this paper,
affecting both the determination of the absolute magnitude
at which the RGB colors are measured and 
the absolute (and then bolometric) magnitude of 
the RGB bump and tip.
Thus our generous estimate of the distance modulus
error yields also conservative estimates of
RGB parameter uncertainties. 
 
It is more difficult to evaluate
the reddening error for each individual cluster.
However, by comparing the values quoted in different catalogs,
we derive an average uncertainty  of a few hundredths  
of magnitude, the largest values 
being those relative to the two most metal rich clusters
in our sample (namely NGC6553 and NGC6528).

\section{The RGB location and morphology: definitions}

Since the main purpose of this paper is the detailed study of the
RGB properties using suitable CMDs, special care is devoted 
to properly define and measure:
{\it i)~} the parameters describing the RGB location in color
and in magnitude
and
{\it ii)~} the parameters describing the RGB morphology, since
they represent the most general and widely used descriptors of
the RGB in any resolved stellar population.

\subsection{The RGB location in color and in magnitude}

Many different observables have been defined and used to locate the
RGB in the classical V,B--V plane (cf. F99 and reference therein),
in the V,V--I plane (Da Costa \& Armandroff 1990) and in the IR planes
(cf. e.g. FCP83).

FCP83 originally defined (V--K)$_0$ (RGB) and (J--K)$_0$ (RGB) as the colors
of the mean RGB ridge line measured at M$_{K}$=--5.5. 
Afterwards, the RGB colors have been measured at different magnitude
levels (M$_K$=--5.0 and M$_K$=--4.0, cf. Cohen \& Sleeper 1995,
hereafter CS95) and the importance of such an approach 
is nicely illustrated by CS95 (cf. e.g. their Figure 3).
In the observational plane, the 
color difference among mean ridge lines for clusters of 
different metallicity becomes progressively more and more pronounced 
ascending the RGB up to the Tip, due to the 
the molecular blanketing effects.

FCP83 also defined the  M$_{K_0}$ (RGB)
as the RGB luminosity at constant color (V--K)$_0$=3.0.
This parameter samples different RGB regions depending from the
cluster metallicity: while in metal--poor
 clusters M$_K$ at (V--K)$_0$=3 marks the
bright upper branch, in the metal--rich ones it intersects the intermediate
RGB, slightly brighter than the HB level. 

To properly map the overall behavior of the RGB as a function 
of the metallicity, in the following we measure the 
(J--K)$_0$ and (V--K)$_0$ colors at different magnitudes:
M$_K$=--3, --4, --5 and --5.5, and the magnitude 
 M$_{K_0}$ (RGB) at fixed color in both the $[M_K,(J-K)_0]$ and
 $[M_K,(V-K)_0]$ planes, 
respectively.

\begin{figure*}[htb]
\vskip6.9truein
\includegraphics{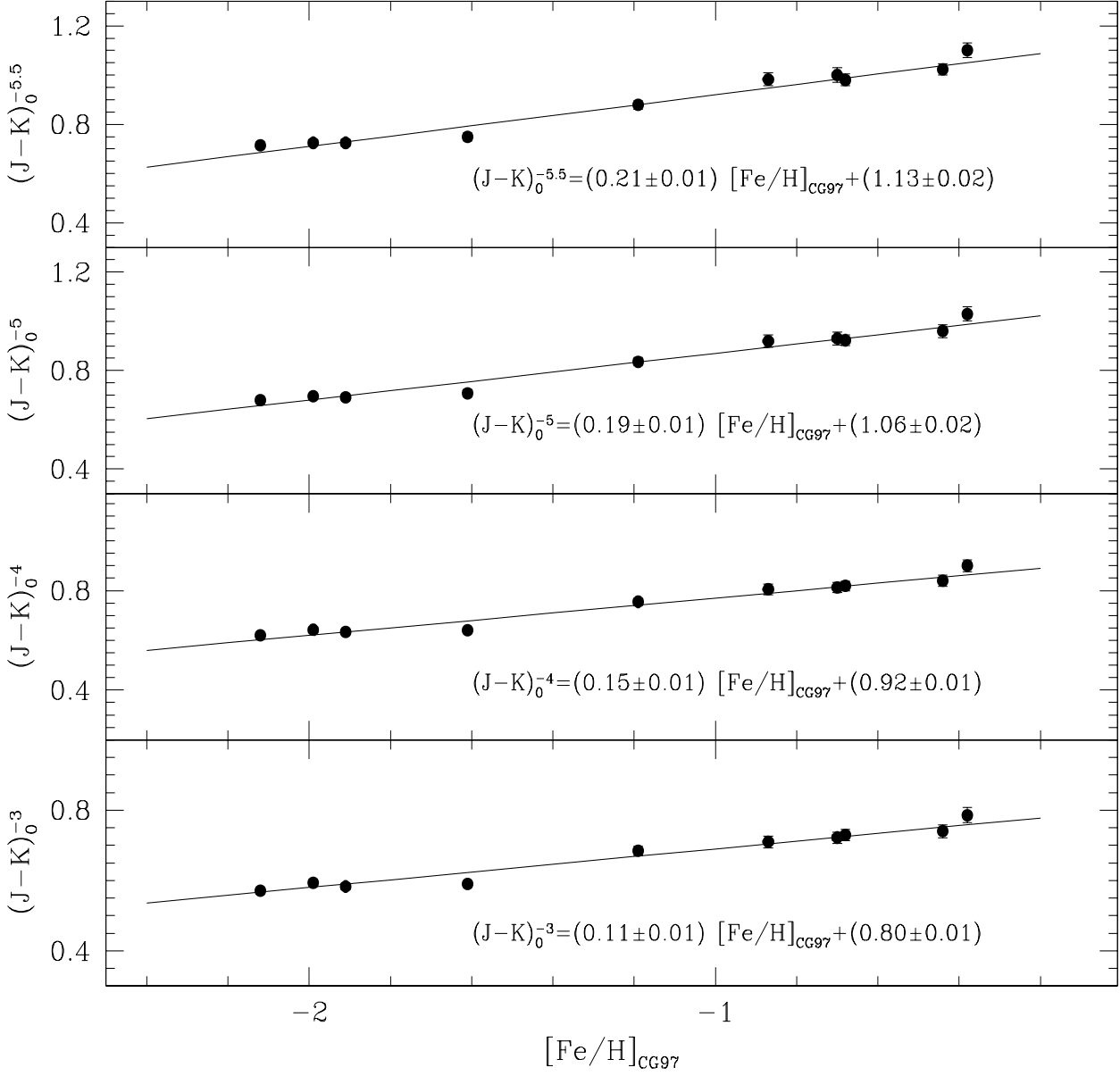}
\caption{
RGB mean (J--K)$_0$ color at different M$_K$ (--3, --4, --5, --5.5)
as a function of CG97 metallicity for the 10 GGCs in our sample.
}
\end{figure*}

\begin{figure*}[htb]
\vskip6.9truein
\includegraphics{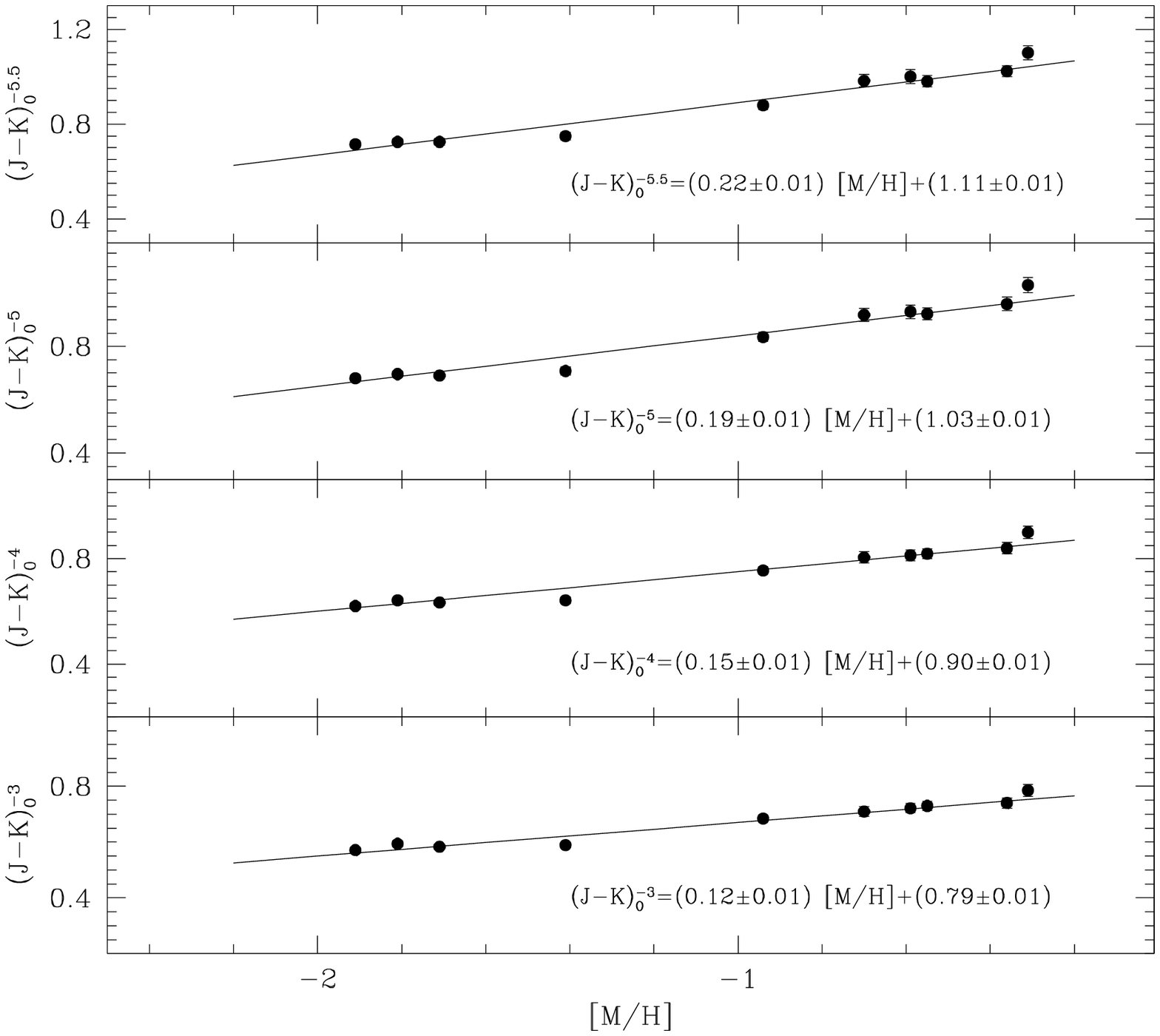}
\caption{
As in Fig.4 but for the {\it global} metallicity scale.
}
\end{figure*}

\begin{figure*}[htb]
\vskip6.4truein
\includegraphics{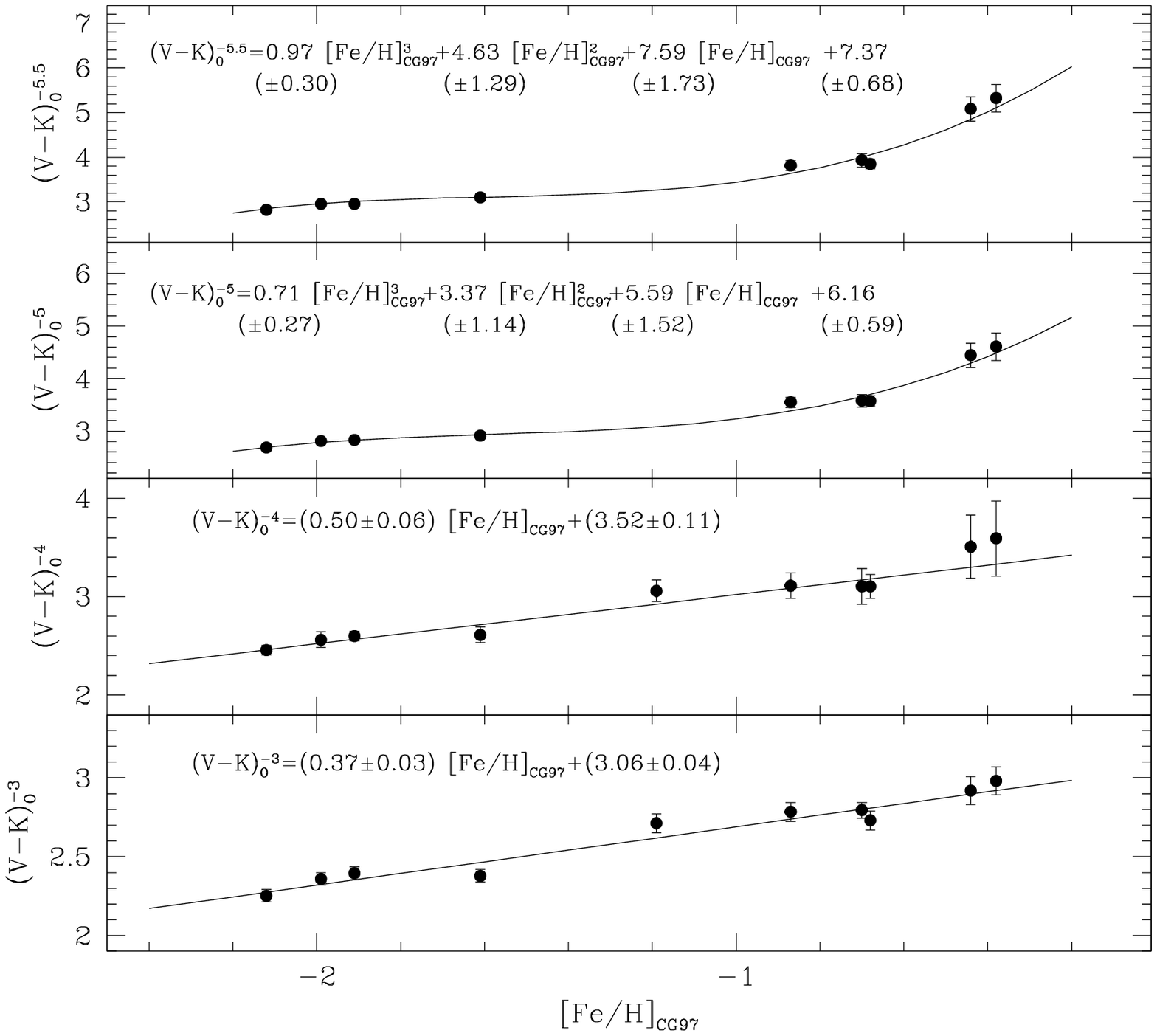}
\caption{
RGB mean (V--K)$_0$ color at different M$_K$ (--3, --4, --5, --5.5)
as a function of CG97 metallicity for the 10 GGCs in our sample.
 }
\end{figure*}

\begin{figure*}[htb]
\vskip6.4truein
\includegraphics{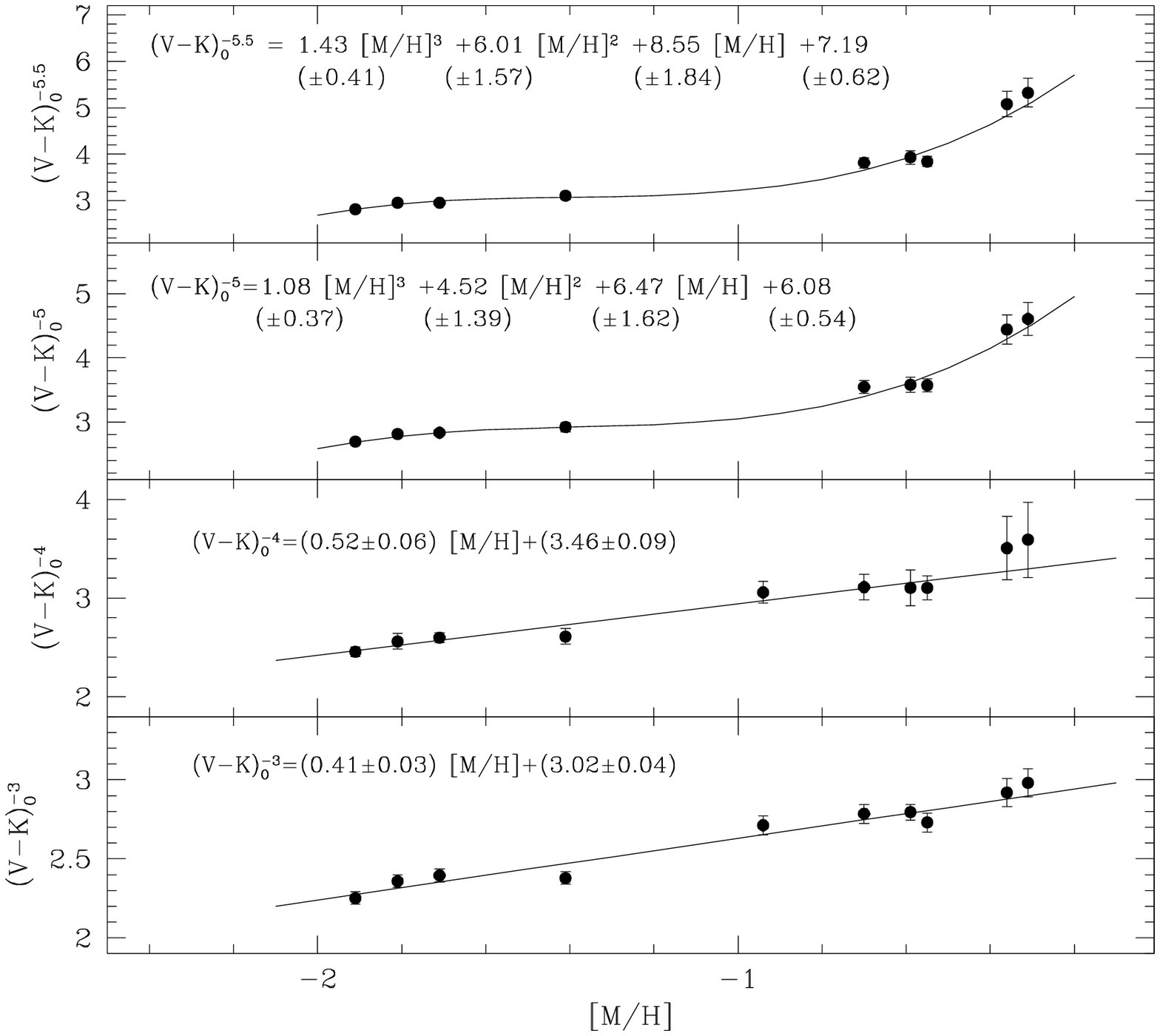}
\caption{
 As in Fig.6 but for the {\it global} metallicity scale.
}
\end{figure*}

\begin{figure*}[htb]
\vskip6truein
\includegraphics{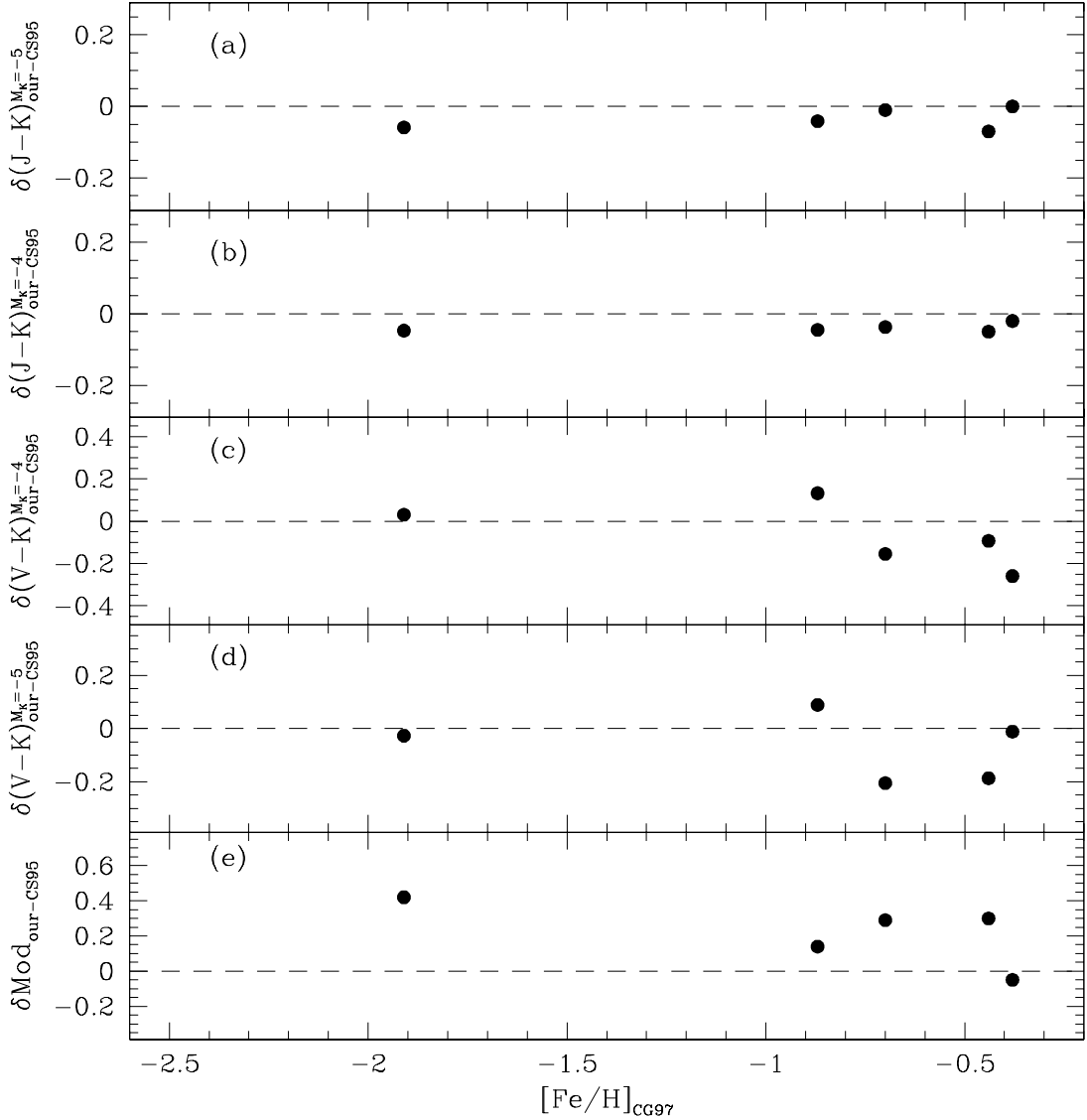}
\caption{
Comparison between our and CS95 RGB
(V-K)$_0^{M_K=-5}$, (V-K)$_0^{M_K=-4}$, (J-K)$_0^{M_K=-4}$,
(J-K)$_0^{M_K=-5}$
colors for the 5 GGCs in common ({\it panels (a),(b),(c),(d)}, respectively).
In {\it panel (e)} we also plotted the difference between  the
 distance moduli adopted here and those adopted by CS95.
 }
\end{figure*}

\begin{figure*}[htb]
\vskip6.3truein
\includegraphics{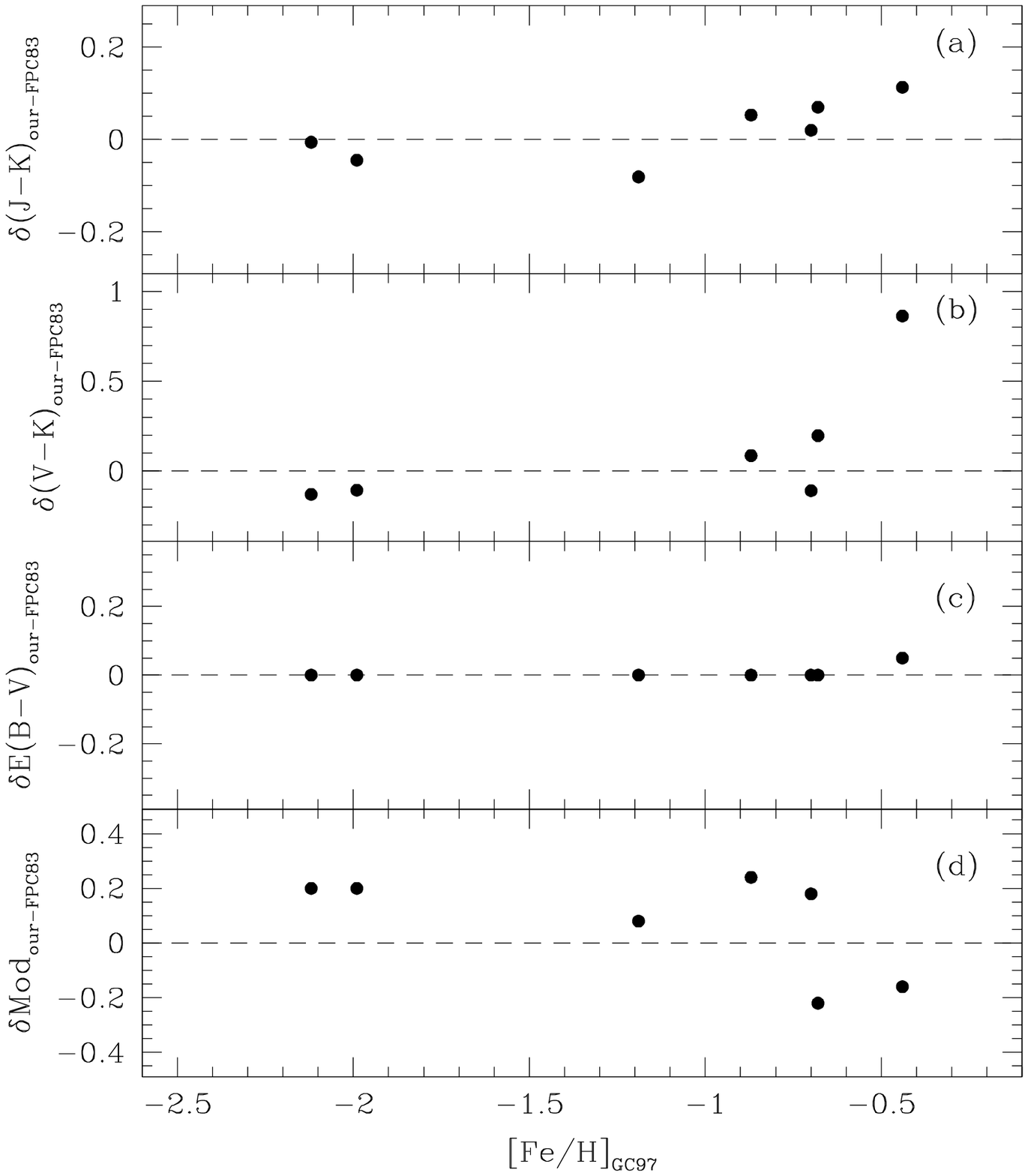}
\caption{
  Comparison between our and FCP83 RGB
(V-K)$_0^{M_K=-5.5}$ and (J-K)$_0^{M_K=-5.5}$ colors
for the 7 GGCs in common ({\it panels (a),(b)}, respectively).
In {\it  panel (c)} and {\it (d)} we  plotted the differences in the
adopted reddening and distance moduli, respectively.
 }
\end{figure*}

\subsection{The RGB slope}

The shape of the RGB in a GGC is sensitive to the abundance of heavy
elements and substantially insensitive to the age and Helium abundance
(cf. Figure 13 by K95).
Moreover, the sensitivity of the RGB slope to the metallicity is highly
pronounced in the observational plane, since an enhancement of the
molecular blanketing strongly affects the RGB colors, making extremely
red the coolest and brightest stars. Consequently, the RGB slope
is a powerful metallicity indicator as shown for example by Ortolani, Barbuy \&
Bica (1991) and K95 (cf. their Figure 15).

K95 presented JHK photometry of 4 high metallicity GGCs (namely NGC5927,  
NGC6712, M71 and Terzan 2). Using this sample and data previously published
for other two clusters (47 Tuc, Frogel et al. 1981; M69, Davidge \& Simons
1991 and FCP83), they derived the RGB slopes in the K,(J--K) plane by
fitting the upper part of the RGB (only the stars between 0.6 and 5.1 mag
brighter than the Zero Age Horizontal Branch (ZAHB) level in K were
considered).

They proposed equations in the form J--K=b+a$\times$K (cf. their Table 9),
and also correlated the RGB slope with the metallicity, finding the
relation: 
$[Fe/H]=-3.9-24.85 ~slope_{RGB}$.
This relation has been later refined by KF95 who found:
$[Fe/H]=-2.89-23.84 ~slope_{RGB}$.
In the present study we derive similar relations for 
the RGB slope in the K,J--K plane from our 
sample of GGCs.

\section{The RGB features: measurements}

In order to perform a complete characterization of the RGB,
in this section we calibrate the main RGB observables
in the IR-CMDs
(namely colors at fixed magnitudes, magnitudes at fixed colors,
 slope, bump and tip) and discuss
their dependence on metallicity.
In doing this we use 
the distance moduli and reddening corrections listed in Table 1
to  locate the observed RGB fiducial ridge lines in the absolute
M$_K$,(J--K)$_0$ and M$_K$,(V--K)$_0$ planes, as
plotted in Fig.3.

\subsection{The RGB colors at fixed K magnitudes}

The intrinsic (J--K)$_0$ and (V--K)$_0$ colors of the RGB measured at
different absolute K magnitudes (M$_K$=--3,--4,--5,--5.5, cf. Sect.5.1)
for the selected clusters are listed in Table 2 and 3, respectively.
In Fig.4--7, we have plotted the RGB (J--K)$_0$ and (V--K)$_0$
colors as a function of the adopted cluster metallicity
in the CG97 ($[Fe/H]_{CG97}$) and the {\it global} ($[M/H]$)
scales. The best fit relation to the data is also
reported in each panel.

As can be seen from Figure 4 and 5, the RGB 
(J--K)$_0$ color scales linearly with [Fe/H],
independently of the selected RGB height-cut, and the slope increases
progressively towards the RGB tip.

On the other hand, the RGB (V--K)$_0$ color 
significantly deviates from a linear dependence
at high metallicity (see the upper two panels in Figs.6 and 7). 
This behavior fully confirms the effect noted
by CS95, who found that the $\Delta $[Fe/H]/$\Delta $(V--K)$_0$ gradient
decreases with increasing [Fe/H].

In order to estimate the possible impact 
on the inferred color -- metallicity relations, 
by using different samples and different assumptions on
the distance scale and reddening,
we compare our results with the two sets of RGB parameters 
obtained by CS95 and FCP83.

\hoffset = -8mm
\begin{deluxetable}{lccccc}
\footnotesize
\tablewidth{18truecm}
\tablecaption{Inferred RGB (J-K)$_0$ colors for the observed GGCs.}
\tablehead{
\colhead{Cluster}&
\colhead{$[Fe/H]$}&
\colhead{$(J-K)_0^{-5.5}$}&
\colhead{$(J-K)_0^{-5}$}&
\colhead{$(J-K)_0^{-4}$}&
\colhead{$(J-K)_0^{-3}$}
 }
\startdata
 NGC104  & --0.70 & $1.000\pm0.029$ & $0.930\pm0.026$ & $0.813\pm0.021$ & $0.722\pm0.016$\nl
 NGC4590 & --1.99 & $0.725\pm0.012$ & $0.696\pm0.011$ & $0.642\pm0.010$ & $0.594\pm0.009$\nl
 NGC6121 & --1.19 & $0.879\pm0.017$ & $0.836\pm0.017$ & $0.755\pm0.015$ & $0.685\pm0.013$\nl
 NGC6171 & --0.87 & $0.983\pm0.026$ & $0.919\pm0.025$ & $0.805\pm0.021$ & $0.710\pm0.017$\nl
 NGC6528 & --0.38 & $1.101\pm0.029$ & $1.030\pm0.028$ & $0.900\pm0.024$ & $0.786\pm0.021$\nl
 NGC6553 & --0.44 & $1.023\pm0.023$ & $0.960\pm0.026$ & $0.840\pm0.022$ & $0.740\pm0.018$\nl
 NGC6637 & --0.68 & $0.980\pm0.024$ & $0.923\pm0.022$ & $0.819\pm0.019$ & $0.730\pm0.016$\nl
 NGC6809 & --1.61 & $0.749\pm0.017$ & $0.708\pm0.015$ & $0.641\pm0.012$ & $0.590\pm0.009$\nl
 NGC7078 & --2.12 & $0.714\pm0.014$ & $0.680\pm0.013$ & $0.620\pm0.011$ & $0.571\pm0.009$\nl
 NGC7099 & --1.91 & $0.725\pm0.014$ & $0.691\pm0.013$ & $0.633\pm0.011$ & $0.583\pm0.009$\nl
\enddata
\end{deluxetable}

\hoffset = -8mm
\begin{deluxetable}{lccccc}
\footnotesize
\tablewidth{18truecm}
\tablecaption{Inferred RGB (V-K)$_0$ colors for the observed GGCs.}
\tablehead{
\colhead{Cluster}&
\colhead{$[Fe/H]$}&
\colhead{$(V-K)_0^{-5.5}$}&
\colhead{$(V-K)_0^{-5}$}&
\colhead{$(V-K)_0^{-4}$}&
\colhead{$(V-K)_0^{-3}$}
}
\startdata
 NGC104  & --0.70 &  $3.93\pm0.15$ &$3.58\pm0.12$ & $3.11\pm0.18$ & $2.80\pm0.05$ \nl
 NGC4590 & --1.99 &  $2.95\pm0.06$ & $2.81\pm0.05$ & $2.56\pm0.08$ & $2.36\pm0.04$ \nl
 NGC6121 & --1.19 &   --           & $3.48\pm0.09$ & $3.06\pm0.11$ & $2.71\pm0.06$ \nl
 NGC6171 & --0.87 &  $3.82\pm0.11$ & $3.55\pm0.10$ & $3.11\pm0.13$ & $2.78\pm0.06$ \nl
 NGC6528 & --0.38 &  $5.33\pm0.31$ & $4.61\pm0.26$ & $3.59\pm0.38$ & $2.98\pm0.09$ \nl
 NGC6553 & --0.44 &  $5.08\pm0.27$ & $4.44\pm0.23$ & $3.51\pm0.32$ & $2.92\pm0.09$ \nl
 NGC6637 & --0.68 &  $3.85\pm0.11$ & $3.57\pm0.10$ & $3.10\pm0.12$ & $2.73\pm0.06$ \nl
 NGC6809 & --1.61 &  $3.10\pm0.08$ & $2.92\pm0.07$ & $2.61\pm0.08$ & $2.38\pm0.04$ \nl
 NGC7078 & --2.12 &  $2.82\pm0.05$ & $2.69\pm0.05$ & $2.46\pm0.05$ & $2.25\pm0.04$ \nl
 NGC7099 & --1.91 &  $2.96\pm0.05$ & $2.83\pm0.05$ & $2.60\pm0.05$ & $2.39\pm0.04$ \nl
\enddata
\end{deluxetable}

\begin{deluxetable}{lcccc}
\footnotesize
 \tablewidth{14cm}
  \tablecaption{Magnitudes and slopes of the RGB for the observed GGCs.}
\tablehead{
\colhead{Cluster} &
\colhead{$[Fe/H]$} &
\colhead{$M_K^{(J-K)_0=0.7}$} &
\colhead{$M_K^{(V-K)_0=3}$} &
\colhead{$slope_{RGB}$}}
\startdata
 NGC 104 & -0.70 & $-2.71\pm0.28$ & $-3.70\pm0.17$ & $-0.104\pm0.002$\nl
 NGC4590 & -1.99 & $-5.07\pm0.36$ & $-5.65\pm0.18$ & $-0.048\pm0.003$\nl
 NGC6121 & -1.19 & $-3.23\pm0.31$ & $-3.85\pm0.37$ & $-0.074\pm0.009$\nl
 NGC6171 & -0.87 & $-2.88\pm0.26$ & $-3.69\pm0.40$ & $-0.101\pm0.005$\nl
 NGC6528 & -0.38 & $-2.11\pm0.20$ & $-3.04\pm0.60$ & $-0.110\pm0.002$\nl
 NGC6553 & -0.44 & $-2.53\pm0.25$ & $-3.17\pm0.57$ & $-0.095\pm0.002$\nl
 NGC6637 & -0.68 & $-2.59\pm0.33$ & $-3.75\pm0.27$ & $-0.092\pm0.002$\nl
 NGC6809 & -1.61 & $-4.89\pm0.29$ & $-5.23\pm0.30$ & $-0.049\pm0.003$\nl
 NGC7078 & -2.12 & $-5.31\pm0.30$ & $-6.14\pm0.31$ & $-0.047\pm0.001$\nl
 NGC7099 & -1.91 & $-5.13\pm0.31$ & $-5.66\pm0.21$ & $-0.043\pm0.003$\nl
\enddata
\end{deluxetable}

\subsubsection{Comparison with CS95}

CS95 presented JK photometry of five GGCs (NGC5927, NGC6528,
NGC 6624, M107 and M30). They also included in their sample an
additional metal rich cluster (NGC6553) observed by Davidge \& Simons (1994a)
and three reference clusters (47 Tuc, M13 and M92) observed by
Cohen et al. (1978) and Frogel et al. (1981).
The adopted reddening, metallicity and distance are listed in their Table
10 together with the measured (J--K)$_0$ and (V--K)$_0$ colors at M$_K$=--4 and
M$_K$=--5.

Five out of the nine GGCs studied by CS95 (47Tuc, M107, NGC6528, NGC6553
and M30) are in common with our sample.
A direct comparison between the (J--K)$_0$ and (V--K)$_0$ colors
at M$_K$=--4 and M$_K$=--5 is then possible.
The residuals (this paper -- CS95) of the RGB colors
as a function of the cluster metallicity are plotted in Fig.8
({\it panel (a), (b), (c), (d)} respectively).

As can be seen, our (J--K)$_0$ colors seem to be systematically $\sim$0.05
mag bluer than the CS95 ones (cf. {\it panels (a)} and {\it (b)} in Fig.8).
We note that the
reddenings assumed here are fully consistent
with those assumed by CS95 (the largest difference being
$\delta E(B-V)\sim 0.06$ mag in the case of NGC6553).
On the other hand, the distance moduli
assumed in this paper are systematically
larger (up to $\sim$0.4 mag,
cf. {\it panel (e)} in Fig.8) than those adopted by CS95.

The slope of the RGB in the M$_K$,(J--K)$_0$
plane is $\sim$--0.1 in metal--rich clusters and $\sim$--0.05 in
metal--poor ones (see Sect.6.3),
hence a difference of --0.2 in the distance modulus 
moves the J--K color at a fixed K magnitude by $\sim$0.01--0.02 mag 
to the red.
Thus, if  we assume the same distance scale as CS95, the discrepancy is
reduced to $\le$0.02 mag, that is well within the level of accuracy of
both photometries.

In summary, the observed discrepancy between our and CS95 (J--K)$_0$ colors
can be mainly ascribed to differences in the assumed distance moduli,
while the reddening and zero--point photometric calibrations are in
reasonable agreement.

Moving to the (M$_K$,(V--K)$_0$) plane, the discrepancy between our 
(V--K)$_0$ colors and those presented by CS95 (cf. {\it panels (c) and (d)} in 
Fig.8) are remarkably larger
(up to about --0.35 mag). Two main factors can be invoked to explain
such a discrepancy.

Different assumptions in the distance moduli produce a variation in the V--K
color which is strongly dependent on the metallicity, due to the
corresponding strong dependence of the RGB slope on the metallicity itself.
Since we assumed larger distance moduli than CS95, we expect to derive
(V--K)$_0$ colors systematically bluer than CS95, with progressively larger
differences at larger metallicity, as indeed observed.

Different calibrations of the zero--point in the V magnitude also
directly affect the (V--K)$_0$ color.
CS95 used old V photometry for NGC6553 (Hartwick 1975),
NGC6528 ( Van Den Berg \& Younger 1979) and NGC6171 (Sandage \& Katem 1964).
While the photometry of NGC6171 by Sandage \& Katem (1964)
is consistent with the one adopted here (Ferraro et al. 1991),
systematic differences are present in the case of NGC6553 and NGC6528.

\subsubsection{Comparison with FCP83}

FCP83 presented RGB parameters for 33 GGCs.
7 out of 10 GGCs presented in this paper are in common with FCP83
(47Tuc, M68, M4, M107, M69, M15 and NGC6553).

The residuals (this paper -- FCP83) for the (J--K)$_0^{-5.5}$ and
(V--K)$_0^{-5.5}$ colors as a function of metallicity are plotted in
Fig.9 ({\it panel (a)} and {\it (b)}, respectively).
From inspecting this figure a possible trend of the (J--K)$_0$ and (V--K)$_0$ 
color residuals as a function of the metallicity is present
(cf. {\it panels (a)} and {\it (b)}), even though based on a
poor statistics to draw any firm conclusions.

Our photometry tends to be bluer in the metal--poor domain
and redder in the metal--rich one. 
While the adopted reddening corrections are almost identical, 
(cf. {\it panel (c)} in Fig.9), it is interesting to note
that, while the distance moduli adopted here are systematically (0.2 mag)
larger than those assumed by FCP83, accounting for possible bluer colors
at a fixed absolute magnitude,
the two most metal rich clusters
among those in common  (namely, NGC6553 and M69)
seem to have distance moduli significantly smaller
than those assumed by FCP83 (cf. {\it panel (d)} in Fig.9).
This fact naturally moves the colors at a fixed magnitude further to the red,
and it is especially effective in the upper part
of the RGB of metal--rich  clusters, for which small variations in magnitude
produce a large variation in color (cf. Fig.3).

Moreover, as discussed in the previous section, the
V photometry zero--point directly affects the (V--K) color and thus
the different optical photometry adopted here
with respect to FCP83 is expected to produce significant differences
especially in the (V--K) color.
In the case of M69, for example, Ferraro et al. (1994a)
showed that the B,V photometry by Hartwick \& Sandage (1968)
and Harris (1977) (adopted by FCP83) has large
(up to $\sim 0.3$ mag) systematic errors.

\subsection{The RGB K magnitudes at fixed colors}

Following FCP83, we measured, from Fig. 3, the K magnitude 
M$_K^{(V-K)_0=3}$ at a fixed (V-K)$_0=3$ color.
In Fig.10 we show the dependence of this parameter
 on metallicity
($[Fe/H]_{CG97}$ in {\it panel (a)} and $[M/H]$ in {\it panel (b)},
respectively). The best fit relation is also reported
in each panel and plotted as a solid line.

It is worth considering
that our linear square fitting  turns out to be significantly
steeper (1.63 and 1.73 using $[Fe/H]_{CG97}$ and $[M/H]$, respectively)
than the FCP83 relation (plotted as a {\it dashed} line in {\it panel (a)}
of Fig.10).
Part of this discrepancy can be ascribed to the adopted metallicity scale.
Using the Zinn scale as FCP83 did, we would in fact obtain a shallower
slope (1.43) than using CG97 metallicity scale, but still
steeper than the FCP83 one ($\sim 1.1$).

In the same vein as for M$_K^{(V-K)_0=3}$, we have defined a similar
parameter in the M$_K$,(J--K)$_0$ plane, namely  M$_K^{(J-K)_0=0.7}$,
the  absolute K magnitude at fixed (J--K)$_0=0.7$ color.
This new parameter allows to characterize the RGB 
from a merely IR CMD and to constrain 
the photometric estimate of metallicity and 
reddening of obscured stellar clusters.

The behavior of this new parameter as a function of metallicity
($[Fe/H]_{CG97}$ in {\it panel (a)} and $[M/H]$ in {\it panel (b)},
respectively) is shown in Fig.11. The best fit solutions are also reported.
As can be seen from the figures, linear relations nicely match the data.

The inferred absolute K magnitude M$_K$ at fixed 
(J--K)$_0$=0.7 and (V--K)$_0$=3 
RGB colors of the selected clusters are listed in 
columns 3 and 4 of Table 4, respectively.
 
The uncertainties in the derived RGB K magnitude
are essentially
driven by the errors in the reddening. As expected,
errors of a few hundredths of magnitude produce uncertainties 
of $\sim 0.2-0.3$ mag in the K magnitudes, depending on the 
RGB region intercepted (see Figure 3). Significative higher
errors are only inferred for the two most metal-rich clusters (NGC6553 
and NGC6528) due to their very uncertain reddening estimates.

\begin{figure*}[htb]
\vskip6.9truein
\includegraphics{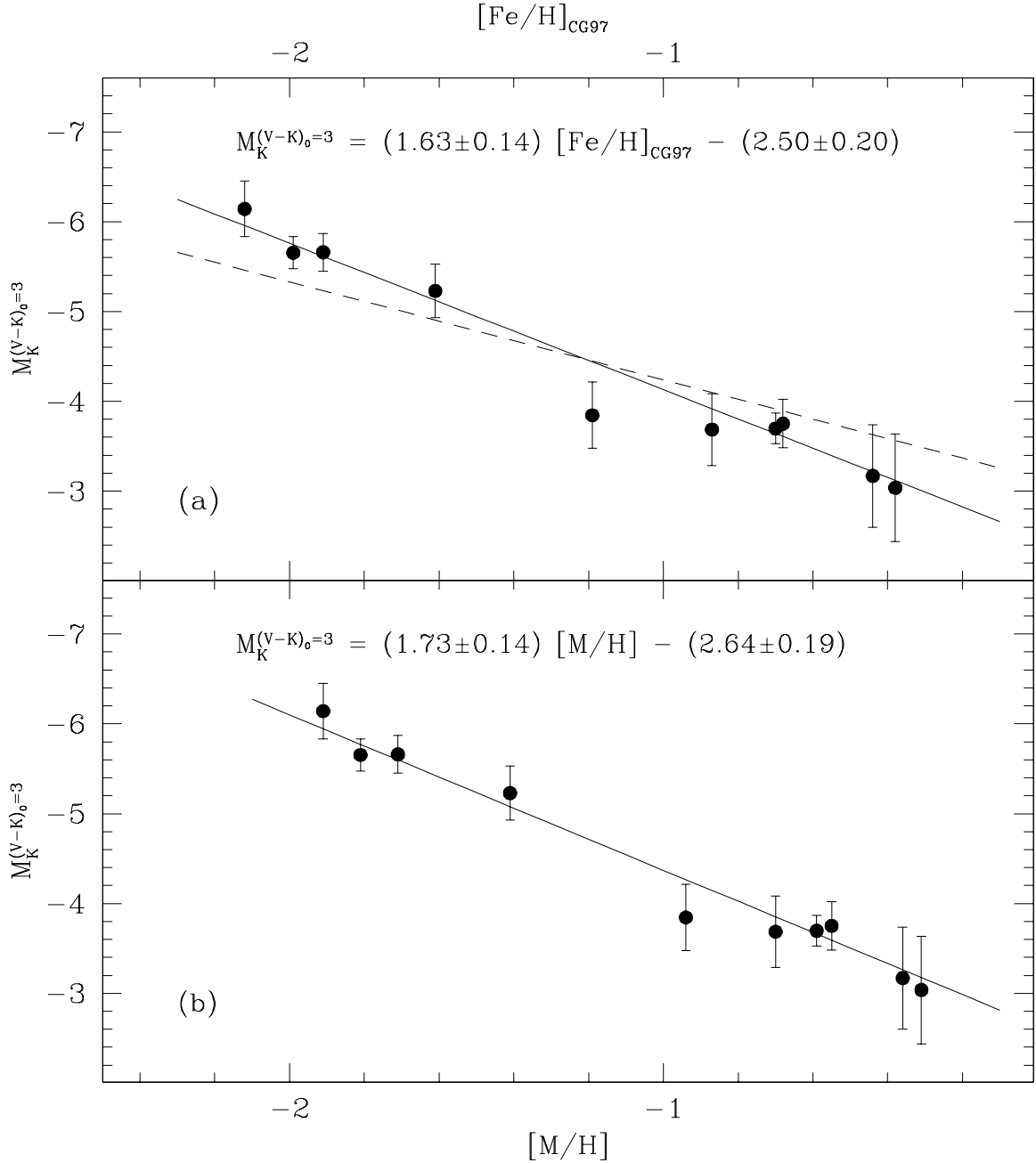}
\caption{
 M$_K$ at constant (V--K)$_0$=3 as a function of the  metallicity
in the CG97 ({\it panel (a)})
and in the {\it global} scale ({\it panel(b)}) , respectively,
for the
10 GGCs in our sample. The solid lines are the best fit to the data, the
dashed line is the FCP83 relation. }
\end{figure*}

\begin{figure*}[htb]
\vskip6.9truein
\includegraphics{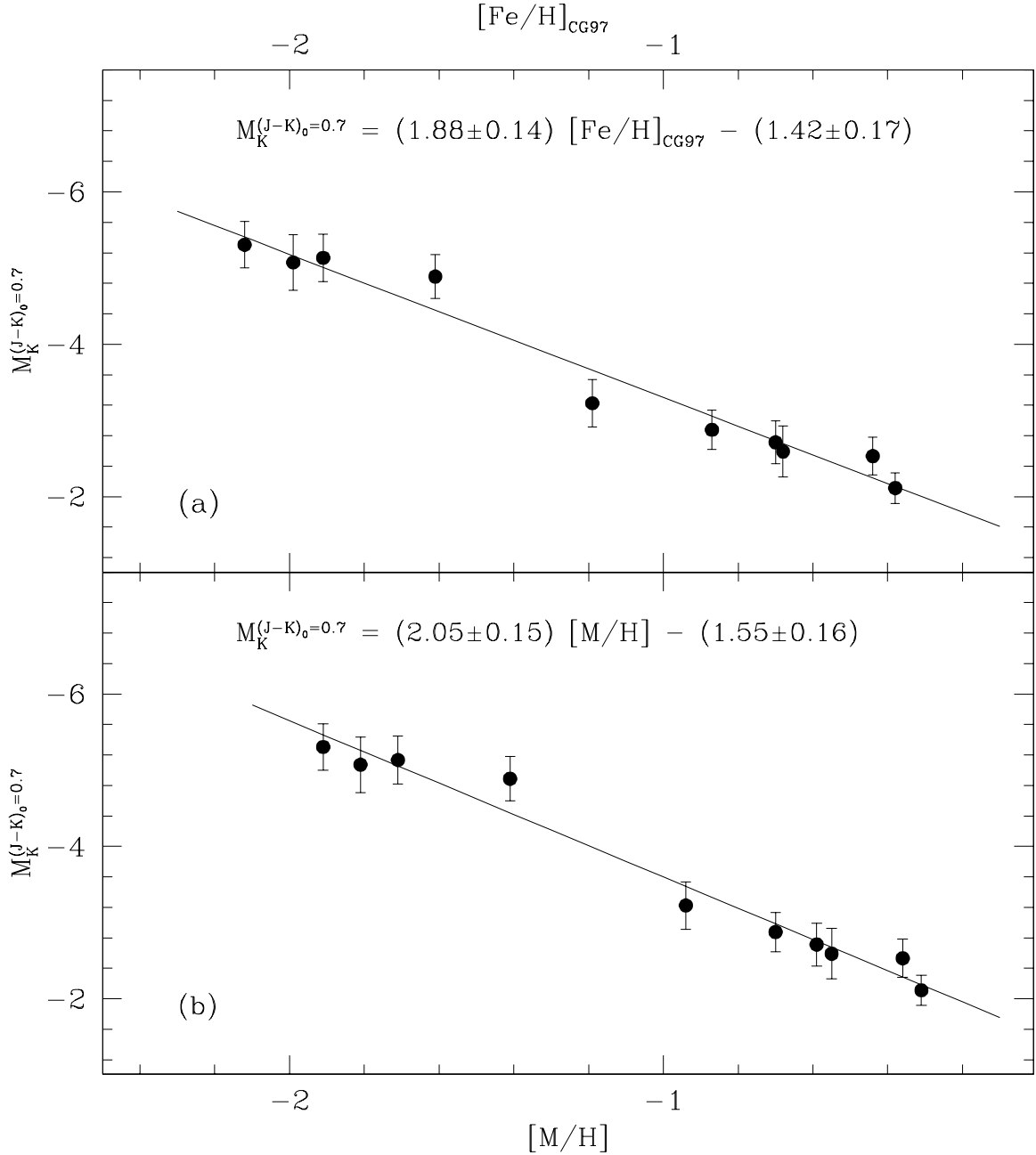}
\caption{
  M$_K$ at constant (J--K)$_0$=0.7 as a function of the  metallicity
in the CG97 ({\it panel (a)})
and in the {\it global} scale ({\it panel(b)}) , respectively,
for the
10 GGCs in our sample. The solid lines are the best fit to the data.
}
\end{figure*}

\begin{figure*}[htb]
\vskip6.9truein
\includegraphics{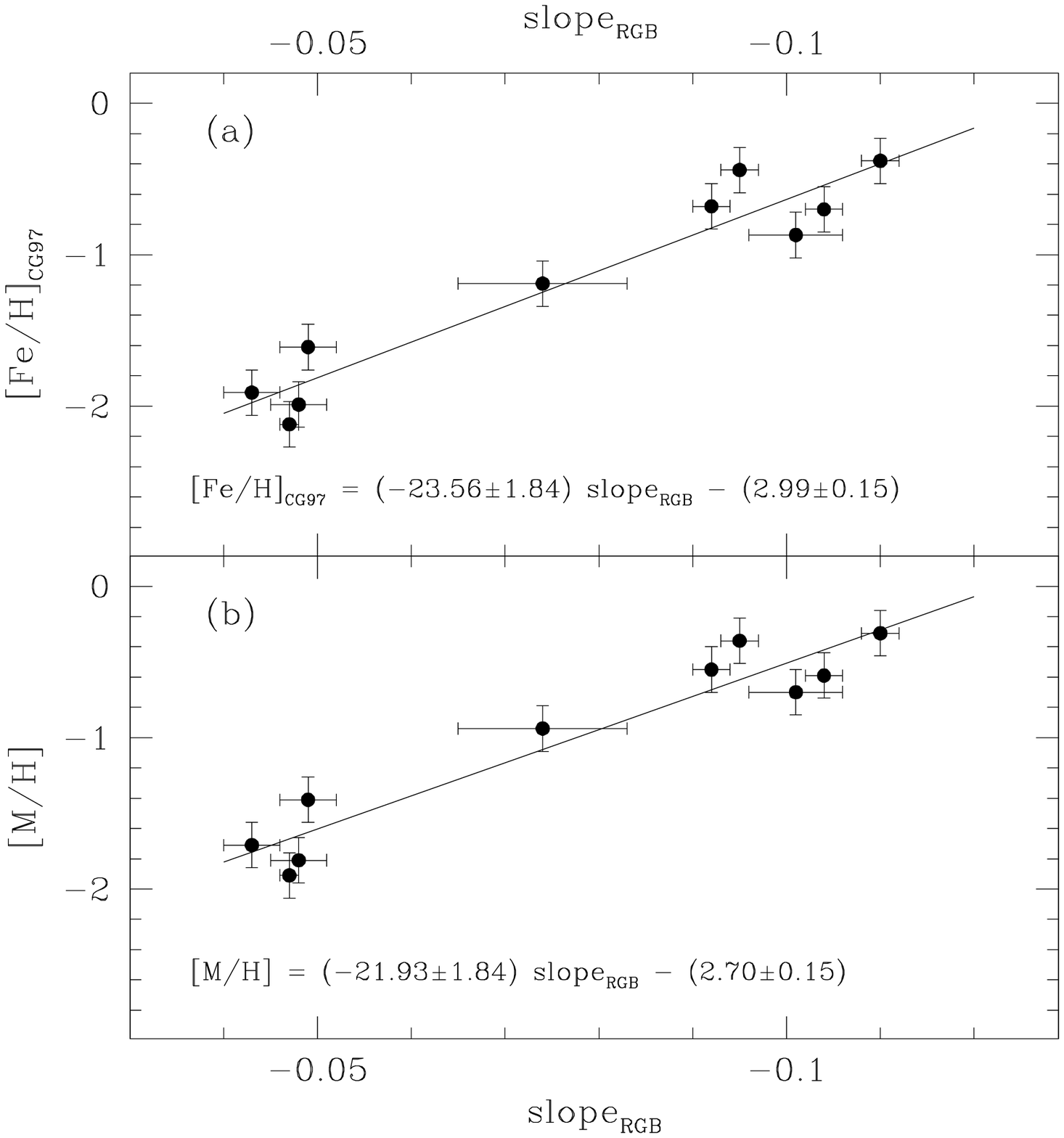}
\caption{
Metallicity scales: $[Fe/H]_{CG97}$ and $[M/H]$, ({\it panel(a)}
and {\it panel (b)}, respectively)  as a
function of the derived RGB slope for the 10 selected GGCs.
}
\end{figure*}

\subsection{The RGB slopes}

To further describe the RGB properties in the IR CMD, we have also
measured the so--called RGB slope ($slope_{RGB}$), adopting the technique
used by K95 and KF95. Even though linear fits are not the best tools
to represent the upper RGB (also in the (K,J--K) plane), they can be used as
a first--order description of the overall RGB morphology.

A basic step before measuring the slope is
to isolate the RGB stars from field objects and members of different
branches. This is not an easy task and we used the RGB samples
here selected as described in Sect.3, where the AGB and HB stars have been
{\it statistically} removed from the CMD, before performing the linear
fit to the data.

The derived slopes for the 10 GGCs in our sample are listed in 
column 5 of Table 4.
From the examination of these values it is evident how 
the RGB slope is a sensitive indicator of the cluster metal content:
the average slope for the 5 most metal rich clusters
in our sample (with $[Fe/H]_{CG97}>-1$) turns out to be
$<slope_{RGB}>=-0.1004\pm0.007$, while it is significantly flatter
($<slope_{RGB}>=-0.047\pm0.003$) in the 4 most metal poor
ones with $[Fe/H]_{CG97}<-1.5$.
This result fully confirms the KF95 suggestion that
the RGB slope in the K,J--K plane is indeed one of the most robust
(reddening--independent) indicator of the cluster metallicity.

Fig.12 shows a linear correlation of the RGB slope with the metallicity
($[Fe/H]_{CG97}$ in {\it panel (a)} and $[M/H]$ in {\it panel (b)},
respectively).
The inferred relations (also reported in Fig.12) are 
fully compatible with those
 by K95 and KF95.
Small differences  can be 
ascribed both to the different metallicity scale adopted here
(CG97 rather than Z85 used by K95 and KF95), and to the fact that their
relations have been derived only for metal rich clusters ($[Fe/H]>-1$),
while those reported in Fig.12 have been derived including also metal--poor
ones. Two GGCs are in common between our and the K95--KF95 samples,
namely 47Tuc and M69. These clusters allow us to perform a direct comparison.
In the case of 47 Tuc  slightly different slopes have been measured:
we get $slope_{RGB}=-0.104$ while K95 found --0.097. 
In the case of  M69 the values are fully in agreement:
we infer $slope_{RGB}=-0.092$ compared to --0.093 by K95. 
However,
we note that the quality (both in terms of photometric accuracy and
sample size) of the  IR CMDs presented here for 47 Tuc and for
the other cluster is superior to the K95 data--base (compare, for
example, the diagrams in Fig.1 with the last two panels of Figure 11 in K95).

\begin{figure*}[htb]
\vskip6.7truein
\includegraphics{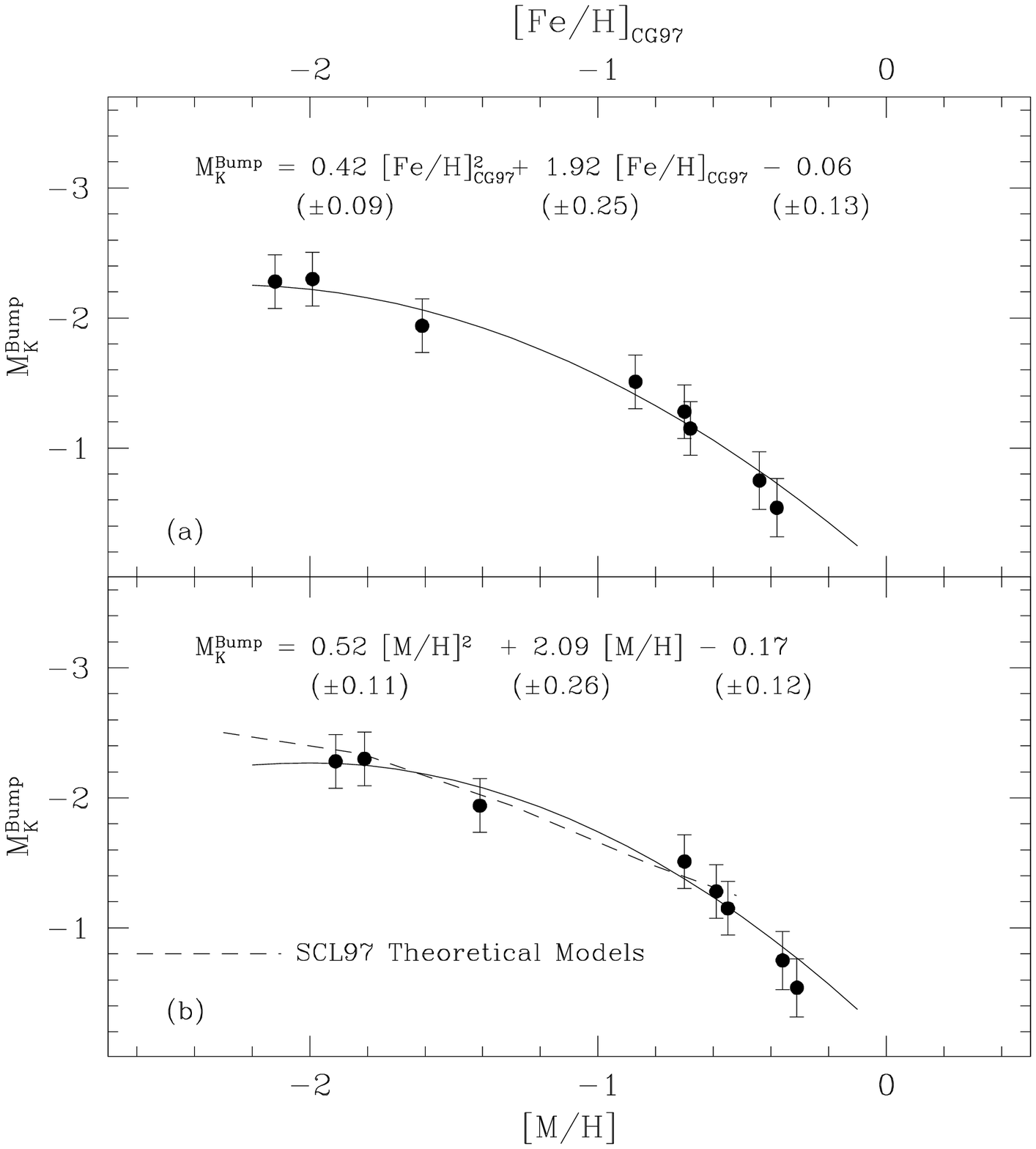}
\caption{
Absolute K magnitude at the RGB bump as a function of
the metallicity (in the CG97 and {\it global}
metallicity scales -- {\it panel(a) and (b)}, respectively),
for the 8 GGCs in our sample, in which the RGB--bump
has been identified. The solid lines are the best fit to the data.
The dashed line in {\it panel (b)} is
the theoretical prediction by SCL97 models at $t=16$ Gyr.
}
\end{figure*}

\subsection{The RGB bump}

One of the most interesting features along the RGB is the so--called
RGB bump (Iben 1968). Its name is due to the peaked distribution of the
differential luminosity function along the RGB.

The RGB evolution is characterized by a narrow burning hydrogen shell
which is moving towards the outer region of the star.
The shell is quite thin in mass and a temporary drop in luminosity is
expected when it reaches the discontinuity in the hydrogen distribution
profile generated by the inner penetration of the convective envelope.
This interruption in the expansion of the stellar envelope has its signature
in the differential LF star excess, the so--called bump.

Fusi Pecci et al. (1990) and recently F99 showed how this
feature can be safely identified in most of the current generation of 
optical CMDs. As already noted by Crocker \& Rood (1984) and Fusi Pecci
et al. (1990), the main difficulty in detecting the RGB bump is having
sufficiently large observational samples (about 2000 RGB stars in
the upper 3 mag).
Moreover, since
in metal--poor clusters the RGB bump
occurs at brighter luminosities, that is
in a region which is intrinsically poorly populated
(towards the upper RGB), its identification is even more difficult.

Following Fusi Pecci et al. (1990) we used the integral and differential LFs
to correctly locate the bump.
In all but two clusters, namely M4 and M30, we succeeded in identifying
the RGB bump with good accuracy.
This is so far the largest near IR sample of Population II stars ever
observed for which it is possible to measure the RGB bump.
The observed K$^{Bump}$ magnitude, the absolute magnitude M$_K^{Bump}$
and the (J--K)$_0$ and (V--K)$_0$ colors of the
RGB bump for the selected clusters are listed in Table 5.
As can be seen, the size of the IR samples
presented here allows a very accurate determination of this feature, 
similar to that one typically obtained from optical CMDs (cf. e.g. 
Table 5 in F99).

In Fig.13 the absolute K magnitude of the RGB bump as a function of the
cluster metallicity in both scales ($[Fe/H]_{CG97} $ and $[M/H]$)
is plotted. The best fit relation to the data
is also plotted as a solid line  
 and the equation reported in each panel. 
As already noted in Sect.4.3, the errors in the 
determinations of the absolute K magnitudes
are mainly driven by the uncertainties in the distance moduli.
 
The dashed line in {\it panel (b)} of
Fig.13  represents the theoretical expectations based on
the SCL97 models, for an age of $t=16$ Gyr (Straniero 1999, private
communication). As can be seen, the models show an excellent agreement
with the observational data.
This result fully confirms the finding of F99
(from the location of the RGB bump in 47 GGCs in the visual band)
that the earlier discrepancy between theory and observations ($\sim 0.4$mag)
(cf. Fusi Pecci et al. 1990, Ferraro 1992) has been completely removed
using the latest theoretical models and the {\it global} metallicity
($[M/H]$).

\begin{figure*}[htb]
\vskip6.4truein
\includegraphics{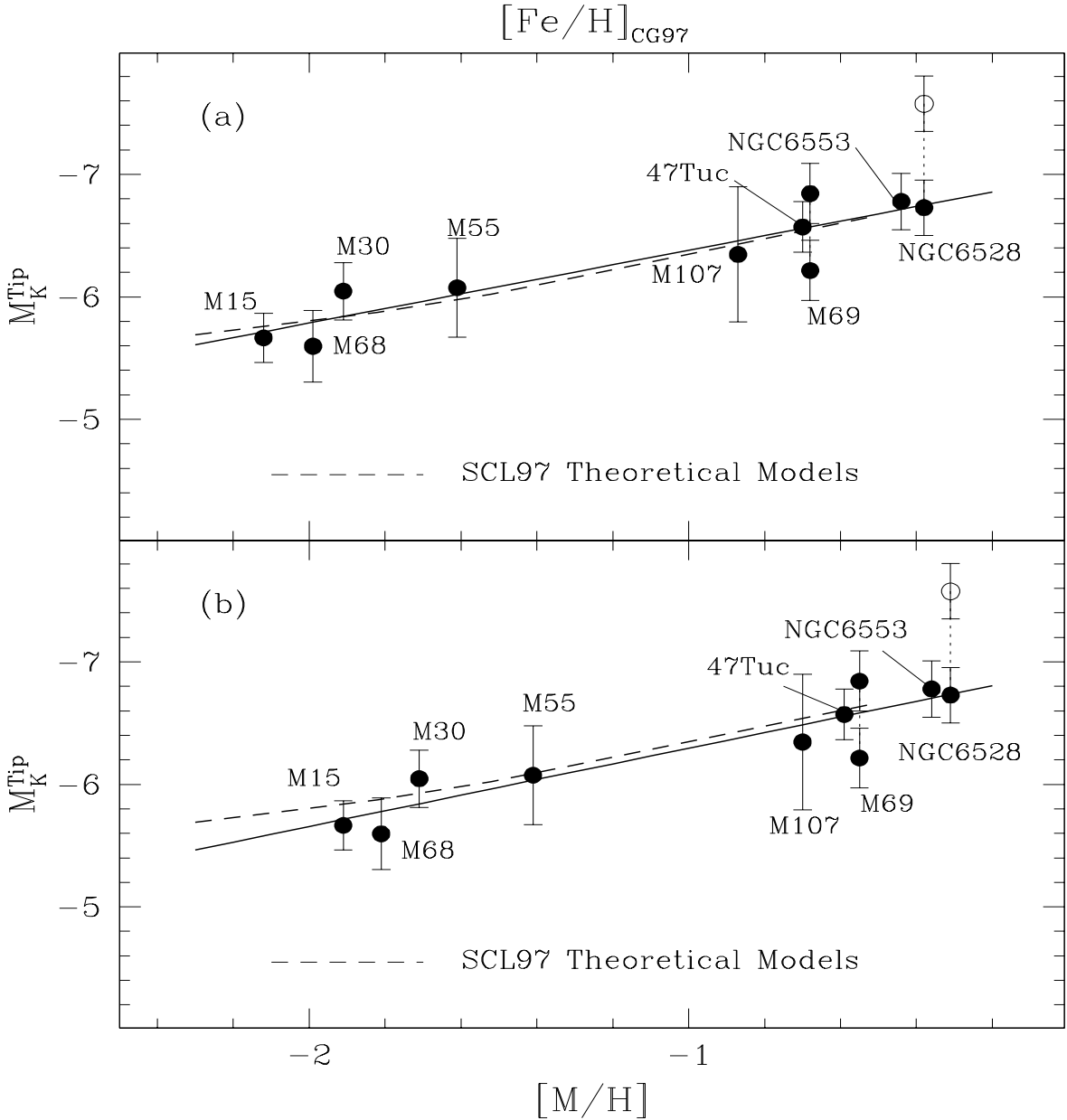}
\caption{
Absolute K magnitude of the brightest star (the RGB tip) as a 
function of the metallicity in the CG97 and {\it global}
scale -- {\it panel(a) and (b)}, respectively --
for the  GGCs in our sample.
Two points have been plotted for NGC6528 and M69
(cf. Sect.6.5 for discussion).
Stars belonging to the same cluster are connected by dotted lines.
The solid lines are the best fit to the data, the 
dashed lines are the theoretical expectations
based on SCL97 models at $t=16$ Gyr.
 }
\end{figure*}

\subsection{The RGB tip}

A quite well defined relationship between the bolometric luminosity of
the brightest RGB star in a GGC and its metallicity has been found
by Frogel, Persson \& Cohen (1981) and FCP83. This finding had a
noteworthy impact both on testing the theoretical models and on
the use of the brightest TGB stars as possible distance indicators.

In order to derive a similar relation using our sample of GGCs,
we identified the {\it candidate} brightest giant in each cluster, 
paying particular care in the decontamination of possible field
objects and (especially in the case of metal--rich clusters) of bright
AGB stars and the variables commonly associated to the AGB, the so--called
Long Period Variables (LPV). Of course, the degree of reliability of
the decontamination is hard to quantify due to the
very small number of stars populating the brightest extreme of the
giant branch, possibly affected by severe statistical fluctuations.

Table 6 reports the absolute K and bolometric magnitudes of the adopted
brightest star (the {\it observed} RGB tip) for 9 of the 10 clusters
considered in this paper (M4 was excluded since the number of sampled
giants is too low). Fig.14 shows the absolute K magnitude of the tip
as a function of the  metallicity in the CG97 and {\it global} scales
({\it panel (a) and (b)}), respectively.

The following best fit relations have been derived:
 $$M_K^{Tip} = -(0.59\pm0.11)[Fe/H]_{CG97}-(6.97\pm0.15)$$
 $$M_K^{Tip} = -(0.64\pm0.12)[M/H]-(6.93\pm0.14)$$
 and they are plotted as solid lines in {\it panel (a)} and 
{\it (b)} of Figure 14, respectively.
Two points have been plotted for NGC6528. This is a very metal--rich
cluster and, like for instance 47 Tuc, it is expected to have a number
of bright variable AGB stars, populating the upper part of the RGB.
These stars have not been identified in this cluster yet,
so a clear discrimination is not possible.
For this reason we have plotted the brightest star in our sample as an empty
circle, but since it could possibly be a LPV, we also consider as
{\it candidate} brightest RGB ({\it non variable})
star the reddest star among the brightest 5 stars in our photometry
(filled circle in Fig.14).

The case of M69 is also worth of a brief discussion since the brightest
not variable star in our photometry is star \#785
(cf. Figure 13 by Ferraro et al. 1994a). However, this star seems too blue
in the (K,V--K) CMD to be a real RGB star (cf. panel (b) of Figure 13
in Ferraro et al. 1994a).
For this reason in this cluster we also consider the second brightest star
(namely \#399). Both these stars have been reported in Table 5 and in Fig.14.

Finally, the dashed line plotted in Fig.14 represents the theoretical
expectation based on SCL97 models
at $t=16$  Gyr (Straniero 1999, private communication).
As can be seen the theoretical prediction nicely agrees with the
observations and, though residual contamination and statistical fluctuations
could still affect the sample, it seems quite rewarding the success
of the theory in reproducing the data. It may also be interesting to note
that such an agreement indirectly implies that the adopted
distances and reddening should not be affected by large errors.

\begin{deluxetable}{lccccccc}
\footnotesize
\tablewidth{18truecm}
\tablecaption{Inferred RGB bump  for the observed GGCs.}
\tablehead{
\colhead{Cluster} &
\colhead{$[Fe/H]$} &
\colhead{$K^{Bump}$} &
\colhead{$M_K^{Bump}$} &
\colhead{$(J-K)_0^{Bump}$} &
\colhead{$(V-K)_0^{Bump}$} &
\colhead{$M_{Bol}^{Bump}$} &
\colhead{$Log(T_e^{Bump})$}}
\startdata
 NGC 104 &  -0.70 &  $12.05\pm0.05$ &  $-1.28\pm0.21$ &  $0.61\pm0.01$ &  $2.41\pm0.04$ & $0.78\pm0.21$ &3.6651 \nl
 NGC4590 &  -1.99 &  $12.85\pm0.05$ &  $-2.30\pm0.21$ &  $0.56\pm0.01$ &  $2.24\pm0.03$ & $-0.42\pm0.21$ &3.6917 \nl
 NGC6121 &  -1.19 &  -- &  -- &  -- &  -- &   -- & --\nl
 NGC6171 &  -0.87 &  $12.55\pm0.05$ &  $-1.51\pm0.21$ &  $0.60\pm0.01$ &  $2.44\pm0.04$ & $0.53\pm0.21$ & 3.6677 \nl
 NGC6528 &  -0.38 &  $14.05\pm0.10$ &  $-0.54\pm0.22$ &  $0.58\pm0.01$ &  $2.30\pm0.04$ & $1.45\pm0.22$ & 3.6751 \nl
 NGC6553 &  -0.44 &  $13.00\pm0.10$ &  $-0.75\pm0.22$ &  $0.59\pm0.01$ &  $2.33\pm0.03$ & $1.26\pm0.22$ &3.6722 \nl
 NGC6637 &  -0.68 &  $13.55\pm0.05$ &  $-1.15\pm0.21$ &  $0.62\pm0.01$ &  $2.29\pm0.03$ & $0.92\pm0.21$ &3.6645 \nl
 NGC6809 &  -1.61 &  $11.90\pm0.05$ &  $-1.94\pm0.21$ &  $0.55\pm0.01$ &  $2.19\pm0.03$ & $-0.10\pm0.21$&3.6980 \nl
 NGC7078 &  -2.12 &  $12.90\pm0.05$ &  $-2.28\pm0.21$ &  $0.54\pm0.01$ &  $2.13\pm0.03$ & $-0.45\pm0.21$&3.6998 \nl
 NGC7099 &  -1.91 &  -- &  -- &  -- &  -- &   -- & --\nl
\enddata
\end{deluxetable}

\begin{deluxetable}{lcccc}
\footnotesize
\tablewidth{14truecm}
\tablecaption{The RGB  tip for the observed GGCs.}
\tablehead{
\colhead{Cluster} &
\colhead{$[Fe/H]$} &
\colhead{$(m-M)_0$} &
\colhead{$M_K^{Tip}$} &
\colhead{$M_{Bol}^{Tip}$}} 
 \startdata
 NGC 104 &  -0.70 &  13.32 &   $-6.57\pm0.20$ & $-3.64\pm0.20$  \nl
 NGC  4590 &  -1.99 &  15.14 & $-5.60\pm0.26$ & $-3.29\pm0.26$ \nl
 NGC 6171 &  -0.87 &  13.95 &  $-6.35\pm0.43$ & $-3.52\pm0.43$ \nl
 NGC6528 &  -0.38 &  14.37 &   $-6.73\pm0.22$ & $-3.68\pm0.22$ \nl
 NGC6553 &  -0.44 &  13.46 &   $-6.78\pm0.22$ & $-3.81\pm0.22$ \nl
 NGC 6637 &  -0.68 &  14.64 &  $-6.84\pm0.23$ & $-3.91\pm0.23$ \nl
 NGC 6637 &  -0.68 &  14.64 &  $-6.21\pm0.23$ & $-3.44\pm 0.23$\nl
 NGC 6809 &  -1.61 &  13.82 &  $-6.07\pm0.33$ & $-3.59\pm 0.33$ \nl
 NGC 7078 &  -2.12 &  15.15 &  $-5.67\pm0.20$ & $-3.36\pm 0.20$ \nl
 NGC 7099 &  -1.91 &  14.71 &  $-6.05\pm0.22$ & $-3.80\pm 0.22$ \nl
 \enddata
\end{deluxetable}

\begin{figure*}[htb]
\vskip6.9truein
\includegraphics{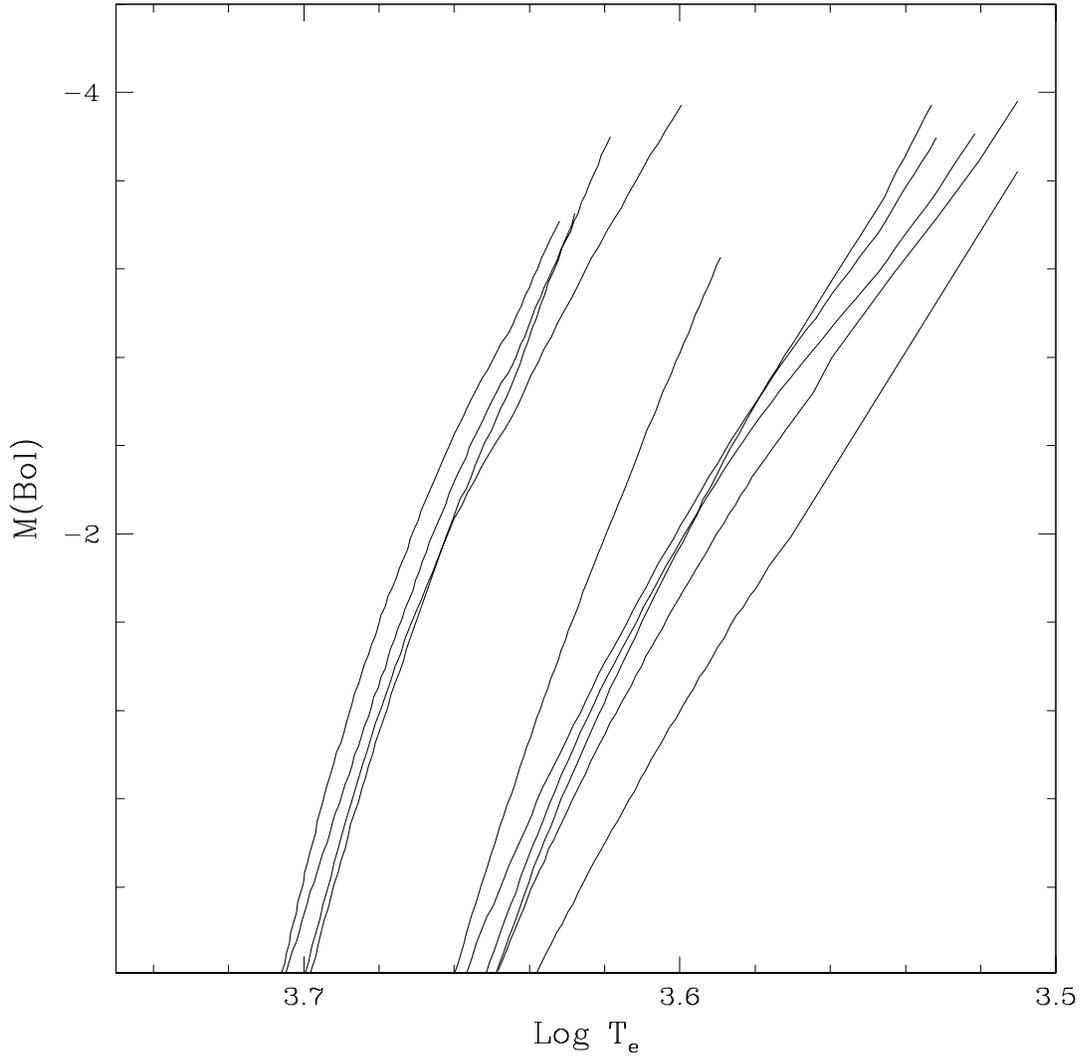}
\caption{
RGB fiducial ridge lines in the (M$_{Bol}$,Log(T$_{e}$))
theoretical plane for the 10 GGCs in our sample.
}
\end{figure*}
 
\begin{figure*}[htb]
\vskip6.9truein
\includegraphics{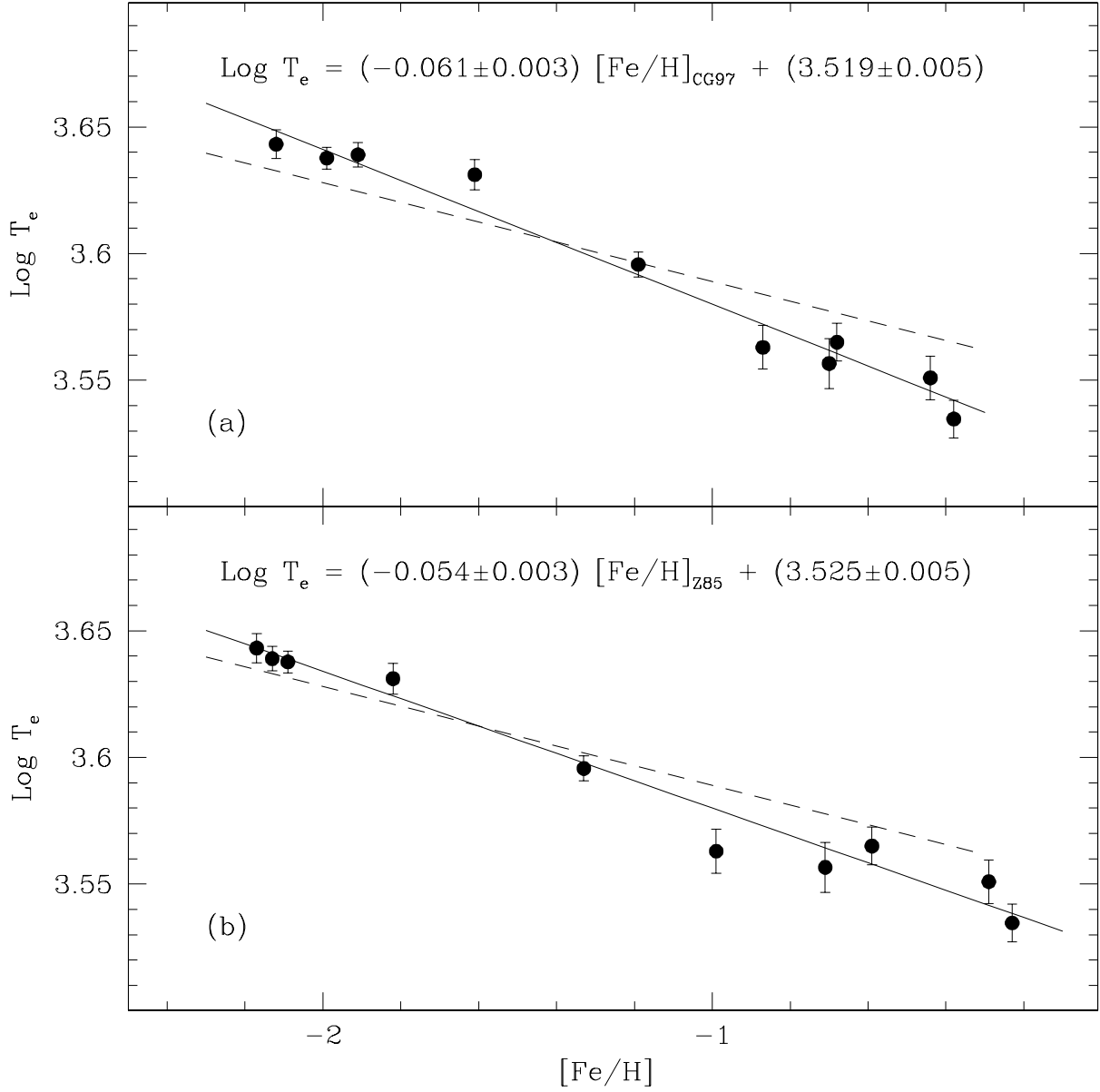}
\caption{
 Log T$_{e}$ at M$_{Bol}$=--3 as a function of
the metallicity (in the CG97 and Z85 scales -
{\it panel(a) and (b)}, respectively) for the 10 GGCs in our sample.
The solid lines are the best fit to the data. The dashed line is the
relation by FCP83.
 }
\end{figure*}

\begin{figure*}[htb]
\vskip6.9truein
\includegraphics{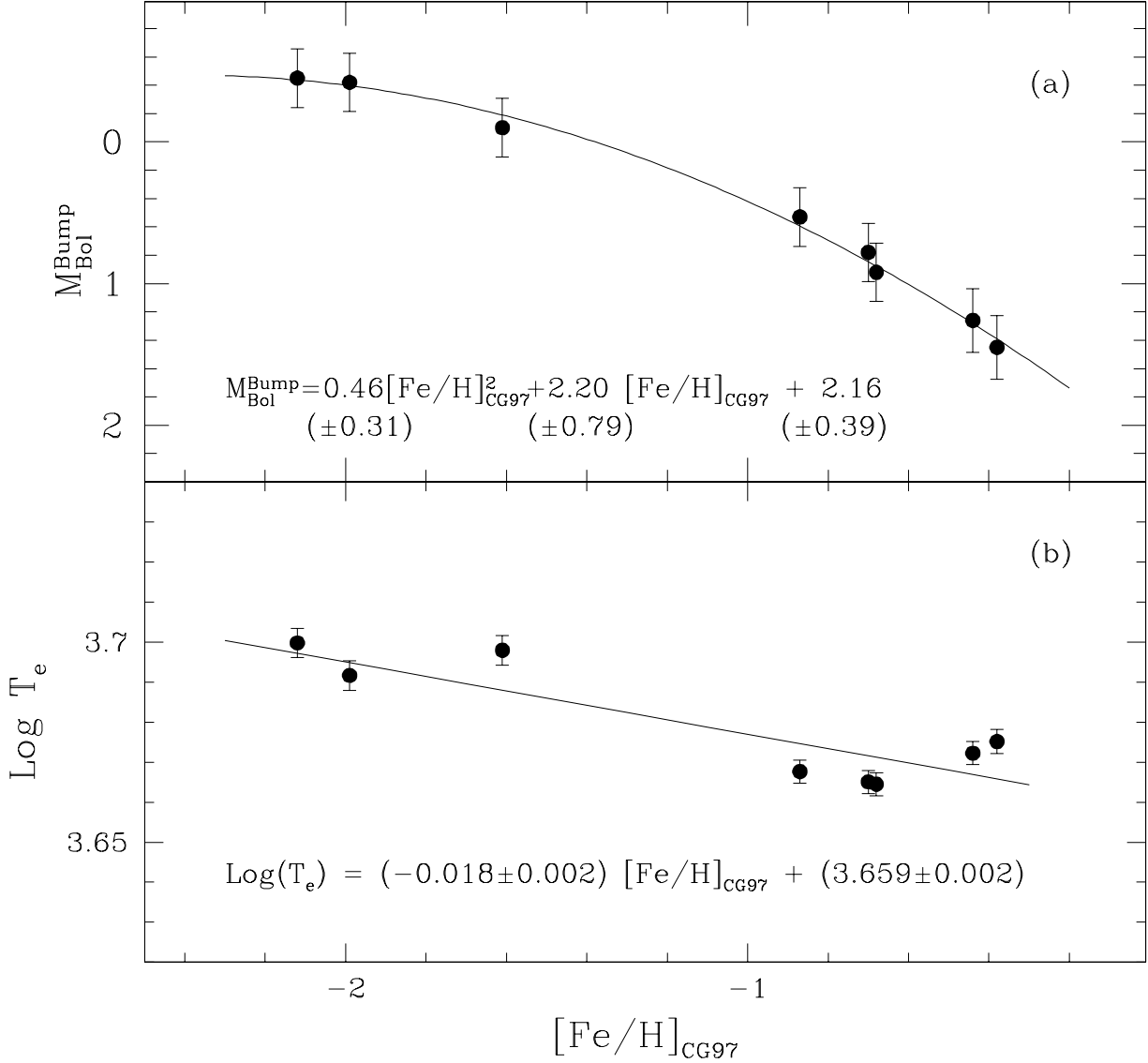}
\caption{
Log T$_{e}$ (bottom panel) and M$_{Bol}$ (upper panel) at the RGB
bump as a function of the CG97 metallicity scale for 8 GGCs in 
our sample.
}
\end{figure*}

\begin{figure*}[htb]
\vskip6.7truein
\includegraphics{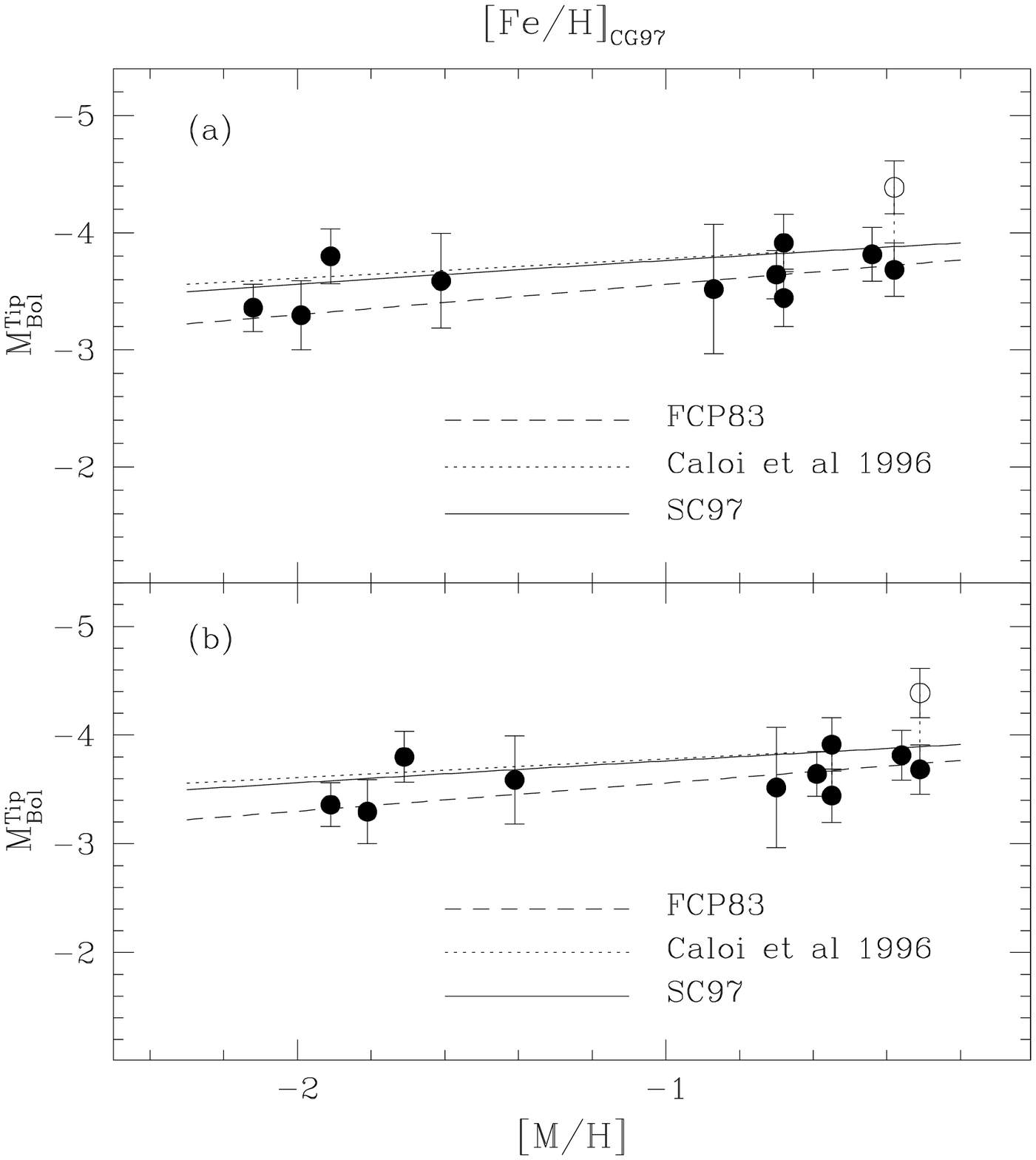}
\caption{
 M$_{Bol}$  of the RGB tip as a function of
metallicity (in the CG97 and {\it global}
scale -- {\it panel(a) and (b)}, respectively)
for 9 GGCs in our sample.
The {\it dashed line} is the relation by FPC83. 
Two theoretical relations have been also plotted: Caloi et al. (1997)[{\it dotted line}] 
and Salaris \& Cassisi (1997, SC97) [{\it solid line}].}
\end{figure*}

\section{The theoretical plane}

The transformations of the observed colors and magnitudes of the RGB features
in the theoretical plane have been performed using the bolometric
corrections and temperature scales for Population II giants
computed and adopted by Montegriffo et al. (1998) (cf. their Table 2).

Since our JK photometry is homogeneous, while the V photometry has been
taken from different  data sets (including HST observations), the
transformations to bolometric magnitudes and effective temperatures were
made in the M$_K$,(J--K)$_0$ rather than in M$_K$,(V--K)$_0$ plane.

In Fig.15 the fiducial RGB ridge lines for the 10 GGCs in our sample
are plotted in the M$_{Bol}$ {\it vs} T$_{e}$ theoretical plane.
From this CMD  we can easily derive the RGB effective temperature at a
fixed bolometric magnitude.
In Fig.16 the effective temperature  at M$_{Bol}$=--3 is shown
as a function of the metallicity.
In particular, adopting the CG97 metallicity scale,  we found the
following relation (cf. {\it panel (a)} in Fig.16):
$$Log T_e = -(0.061\pm0.003)[Fe/H]_{CG97} + (3.519\pm0.005) $$
If the {\it global} metallicity
is adopted, the relation turns out to be slightly different: 
$$ Log T_e = -(0.064\pm0.003)[M/H] + (3.527\pm0.005)  $$
The {\it formal} uncertainties for the derived effective temperatures
turn to be $\sim 50-100 ^{o}K$.

The inferred relations are steeper than in  FCP83 (the most metal--poor
clusters being hotter and the most metal--rich ones cooler
than in FCP83). This effect can be mainly ascribed to the
different
metallicity scales adopted here, as in the case of the the absolute K
magnitude (cf. Fig.8).
In fact, using the Zinn's metallicity scale we find a
shallower relation:
$$Log T_e = -(0.054\pm0.003)[Fe/H]_{Z85} + (3.526\pm0.005)$$
\noindent very similar to the FCP83 one (cf. dashed line in 
{\it panel (b)} of Fig.16).

In Fig.17 we plotted the bolometric magnitude {\it (panel (a))} 
and the effective temperature {\it (panel (b))} of the RGB bump 
as a function of metallicity. 
In {\it panel (a)}, a quadratic
relation has been derived to describe the bolometric magnitude of the
RGB--bump as a function of GC97 metallicity.
The corresponding relation using the {\it global} metallicity
turns to be:
$$ M_{Bol}^{Bump} = (0.64\pm0.37)[M/H]^2+(2.53\pm0.85)[M/H]+(2.10\pm0.35)$$
In {\it panel (b)},
the effective temperature is plotted as a function of GC97 metallicity and
the best fit relation is reported.
If the {\it global} metallicity ($[M/H]$) is
adopted, we get the following  relation:
$$ Log T_e = -(0.020\pm0.002)[M/H] + (3.660\pm0.002)  $$
Finally, Fig.18 reports the bolometric magnitude of the
RGB--tip for the selected clusters as determined in the previous section.
The best fit relations to our data are:
$$M_{Bol}^{Tip}=-(0.25\pm0.11) [Fe/H]_{CG97} - (3.96\pm0.13)$$
and
$$M_{Bol}^{Tip}=-(0.27\pm0.12) [M/H] - (3.94\pm0.13)$$
adopting the two metallicity scales, respectively.
As can be seen, our result is fully consistent with the one obtained
by FPC83 (plotted as {\it dashed line} in Figure 18):
$$M_{Bol}^{Tip}=-0.26 [Fe/H]_{CG97} -3.82$$
However, it is worthy of noticing  that  shallower relations are obtained 
if the brightest RGB star in NGC6528 (plotted as an open circle in Figure 18)
is not considered in the fit:
$$M_{Bol}^{Tip}=-(0.15\pm0.11) [Fe/H]_{CG97} - (3.79\pm0.15)$$
and
$$M_{Bol}^{Tip}=-(0.16\pm0.12) [M/H] - (3.78\pm0.14)$$
adopting the two metallicity scales, respectively.

For further comparisons with the models, two theoretical relations have
been over--plotted in Fig.18: Caloi et al. (1997) (dotted line)
and Salaris  \& Cassisi (1997, SC97) (solid line), respectively.
They nicely agree each other and both  define the upper  boundary of
the observations. It is important to remind here that the theoretical
relationships have to be considered indeed as upper limits to the
luminosity of the observed giants, because of the statistical fluctuations
affecting these poor samples (Castellani, Degl'Innocenti, Luridana, 1993).
On the other hand, it is also worth noticing  that the bolometric
magnitudes obtained from theoretical models are themselves affected by
a quite large and systematic uncertainty ($\ge$ 0.1 mag), depending,
among others,
on the adopted bolometric correction for the Sun (Straniero 1999,
private communication).

\subsection {Uncertainties in the RGB tip determination}

The procedure followed in order to compute the 
errors in the determination of the RGB tip magnitude deserves a
brief description.

As demonstrated by Rood \& Crocker (1997)
 a first order estimation of the statistical error
($\sigma_{stat}$)
in determining the RGB tip from a given sample of RGB stars,
mainly depends on the sample size,  
being $\sigma_{stat}^2 = 1/(\alpha (N+1) (N+2))$,
where N is the number of stars in the upper two bolometric 
magnitudes and $\alpha$ is a parameter which depends
on the rate of the RGB evolution. Its typical value is 0.04.
In our sample of clusters $\sigma_{stat}$
turns to be $\le 0.15$ mag. Only 
two clusters (namely M107 and M55),
 for which less than 10 stars 
in the upper two bolometric magnitudes of the RGB  
have been found,  have larger scatters 
(0.39 and 0.26 mag, respectively).
 Of course, in order to compute the global uncertainty of the absolute 
magnitude, $\sigma_{stat}$
has to be combined with the error in the determination of the distance
modulus ($\sim 0.2$ mag). 
The final errors on the 
bolometric RGB--Tip magnitude are listed
in column  5 of Table 6.
These errors have been conservatively
assumed also for the M$_K^{Tip}$ (see column 4 of Table 6).

\section{Conclusions}    
A new set of high quality IR Color Magnitude Diagrams  has been
presented for a sample of 10 GGCs, spanning a wide range in metallicity.
This new, homogeneous data--base has been used to determine a variety of
observables quantitatively describing the main properties of the Red
Giant Branch, namely:
{\it (a)} the location of the RGB in the CMD (both in (J--K)$_0$ and 
(V--K)$_0$ colors at different absolute K magnitudes 
(--3, --4, --5, --5.5) and in temperature);
{\it (b)} its overall morphology and slope;
{\it (c)} the luminosity of the Bump and the Tip.
All these quantities have been measured via a homogeneous procedure
applied to each individual CMD.
Their behavior as a function of the cluster metallicity has been 
investigated,
also taking into account the effect of the $\alpha$--enhancement.

Comparisons with the most updated theoretical models show a substantial good
agreement between observations and theoretical expectations, taking into
account the errors and uncertainties still affecting both data and
models.

In particular, it is interesting to note that the distance scale adopted
in this paper (from F99), based on the theoretical
luminosity of the ZAHB level, yields a very good agreement between the
observed features and the theoretical expectations along the RGB.
If confirmed, this would indicate a high level of self--consistency
of the theoretical RGB and ZAHB models, computed with the most
up--dated input physics.

The relationships we present here can be used in further
studies {\it i)} to derive a {\it photometric} estimate of the metal 
abundance from
the RGB morphology and location, {\it ii)} to get useful distance
estimates using both the RGB--tip and RGB--bump luminosities, and {\it iii)} 
to describe the overall behavior of the RGB properties in the
near IR. 

By coupling this IR study with 
our previous one on 61 GGCs in the V,B--V
plane (cf. F99), a quite global scenario emerges, where 
the observational and theoretical descriptions start to match quite 
nicely if suitable multi--band photometric samples and 
up--dated models are used.

\acknowledgments

We warmly thank Oscar Straniero for making available the M$_K$ magnitude of
the RGB bump and tip from SCL97 models, and Bob Rood
for the many instructive
and useful discussions on theoretical models.
The financial support of the {\it ``Ministero della Universit\`a e della Ricerca
Scientifica e Tecnologica''} (MURST) to the project {\it Stellar Evolution}
is kindly acknowledged. FRF acknowledges the {\it ESO Visitor Program}
for the hospitality.

\end{document}